

\font\eightrm=cmr8 at 8pt
\font\tenrm=cmr12 at 10pt

\font\seventeenrm=cmr17 at 17pt
\font\twentyonerm=cmr17 at 21pt

\font\ss=cmss10

\font\twelvecal=cmsy10 at 12pt

\font\twelvemath=cmmi12

\font\eightboldmath=cmmib10 at 8pt

\font\seventeenbold=cmbx7 at 17pt

\font\fively=lasy5
\font\sevenly=lasy7
\font\tenly=lasy10

\textfont10=\tenly
\scriptfont10=\sevenly
\scriptscriptfont10=\fively
\magnification=1200
\parskip=10pt
\parindent=20pt
\def\today{\ifcase\month\or January\or February\or March\or April\or May\or
June
       \or July\or August\or September\or October\or November\or December\fi
       \space\number\day, \number\year}

\def\title#1{\footline={\ifnum\pageno<2\hfil
       \else\hss\tenrm\folio\hss\fi}\vskip1truein\centerline{{#1}
       \footnote{\raise1ex\hbox{*}}{\eightrm Supported in part
       by the Robert A. Welch Foundation and N.S.F. Grant
       PHY9009850.}}}

\def\Z{\hfill\break}
\def\newpage{\vfill\eject}
\def\abstract#1{\centerline{\bf ABSTRACT}\vskip.2truein{\narrower\noindent#1
       \smallskip}}
\def\acknowledgements{\noindent\line{\bf Acknowledgements\hfill}\nobreak
    \vskip.1truein\nobreak\noindent\ignorespaces}
\def\runninghead#1#2{\voffset=2\baselineskip\nopagenumbers
       \headline={\ifodd\pageno\rightheadline\else \leftheadline\fi}
       \def\rightheadline{{\sl#1}\hfill{\rm\folio}}
       \def\leftheadline{{\rm\folio}\hfill{\sl#2}}}

\newcount\footnoteno
\def\Footnote#1{\advance\footnoteno by 1
                \let\SF=\empty
                \ifhmode\edef\SF{\spacefactor=\the\spacefactor}\/\fi
                $^{\the\footnoteno}$\ignorespaces
                \SF\vfootnote{$^{\the\footnoteno}$}{#1}}

\def\place#1#2#3{\vbox to0pt{\kern-\parskip\kern-7pt
                             \kern-#2truein\hbox{\kern#1truein #3}
                             \vss}\nointerlineskip}
\def\figurecaption#1#2{\kern.75truein\vbox{\hsize=5truein\noindent{\bf Figure
    \figlabel{#1}:} #2}}
\def\tablecaption#1#2{\kern.75truein\lower12truept\hbox{\vbox{\hsize=5truein
    \noindent{\bf Table\hskip5truept\tablabel{#1}:} #2}}}
\def\boxed#1{\lower3pt\hbox{
                       \vbox{\hrule\hbox{\vrule

\vbox{\kern2pt\hbox{\kern3pt#1\kern3pt}\kern3pt}\vrule}
                         \hrule}}}
\def\a{\alpha}
\def\b{\beta}
\def\g{\gamma}\def\G{\Gamma}
\def\d{\delta}\def\D{\Delta}
\def\e{\epsilon}
\def\z{\zeta}

\def\k{\kappa}
\def\l{\lambda}
\def\m{\mu}
\def\n{\nu}

\def\p{\pi}\def\P{\Pi}\def\vp{\varpi}
\def\r{\rho}

\def\t{\tau}

\def\ph{\phi}
\def\ch{\chi}

\def\o{\omega}\def\O{\Omega}

\def\ca#1{\relax\ifmmode {{\cal #1}}\else $\cal #1$\fi}

\def\calb{{\cal B}}

\def\calm{{\cal M}}

\def\inbar{\vrule height1.5ex width.4pt depth0pt}
\def\IB{\relax{\rm I\kern-.18em B}}
\def\IC{\relax\hbox{\kern.25em$\inbar\kern-.3em{\rm C}$}}
\def\ID{\relax{\rm I\kern-.18em D}}
\def\IE{\relax{\rm I\kern-.18em E}}
\def\IF{\relax{\rm I\kern-.18em F}}
\def\IG{\relax\hbox{\kern.25em$\inbar\kern-.3em{\rm G}$}}
\def\IH{\relax{\rm I\kern-.18em H}}
\def\II{\relax{\rm I\kern-.18em I}}
\def\IK{\relax{\rm I\kern-.18em K}}
\def\IL{\relax{\rm I\kern-.18em L}}
\def\IM{\relax{\rm I\kern-.18em M}}
\def\IN{\relax{\rm I\kern-.18em N}}
\def\IO{\relax\hbox{\kern.25em$\inbar\kern-.3em{\rm O}$}}
\def\IP{\relax{\rm I\kern-.18em P}}
\def\IQ{\relax\hbox{\kern.25em$\inbar\kern-.3em{\rm Q}$}}
\def\IR{\relax{\rm I\kern-.18em R}}
\def\IZ{\relax\ifmmode\hbox{\ss Z\kern-.4em Z}\else{\ss Z\kern-.4em Z}\fi}
\def\IGa{\relax{\rm I}\kern-.18em\Gamma}
\def\IPi{\relax{\rm I}\kern-.18em\Pi}
\def\ITh{\relax\hbox{\kern.25em$\inbar\kern-.3em\Theta$}}
\def\IOm{\relax\thinspace\inbar\kern1.95pt\inbar\kern-5.525pt\Omega}


\def\ie{{\it i.e.,\ \/}}

\def\noblackboxes{\overfullrule=0pt}
\def\define{\buildrel\rm def\over =}

\def\cym{Calabi--Yau manifold}
\def\cys{Calabi--Yau manifolds}

\def\K{K\"ahler}

\def\H#1#2{\relax\ifmmode {H^{#1#2}}\else $H^{#1 #2}$\fi}
\def\M{\relax\ifmmode{\calm}\else $\calm$\fi}

\def\Bigcheck{\lower3.8pt\hbox{\smash{\hbox{{\twentyonerm \v{}}}}}}
\def\bigboldcheck{\smash{\hbox{{\seventeenbold\v{}}}}}

\def\Bighat{\lower3.8pt\hbox{\smash{\hbox{{\twentyonerm \^{}}}}}}

\def\Msharp{\relax\ifmmode{\calm^\sharp}\else $\smash{\calm^\sharp}$\fi}
\def\Mflat{\relax\ifmmode{\calm^\flat}\else $\smash{\calm^\flat}$\fi}
\def\preMcheck{\kern2pt\hbox{\Bigcheck\kern-12pt{$\cal M$}}}
\def\Mcheck{\relax\ifmmode\preMcheck\else $\preMcheck$\fi}
\def\preMhat{\kern2pt\hbox{\Bighat\kern-12pt{$\cal M$}}}
\def\Mhat{\relax\ifmmode\preMhat\else $\preMhat$\fi}

\def\Bsharp{\relax\ifmmode{\calb^\sharp}\else $\calb^\sharp$\fi}
\def\Bflat{\relax\ifmmode{\calb^\flat}\else $\calb^\flat$ \fi}
\def\preBcheck{\hbox{\Bigcheck\kern-9pt{$\cal B$}}}
\def\Bcheck{\relax\ifmmode\preBcheck\else $\preBcheck$\fi}
\def\preBhat{\hbox{\Bighat\kern-9pt{$\cal B$}}}
\def\Bhat{\relax\ifmmode\preBhat\else $\preBhat$\fi}

\def\figBcheck{\kern3pt\hbox{\raise1pt\hbox{\bigboldcheck}\kern-11pt
    {\twelvecal B}}}
\def\figBsharp{{\twelvecal B}\raise5pt\hbox{$\twelvemath\sharp$}}
\def\figBflat{{\twelvecal B}\raise5pt\hbox{$\twelvemath\flat$}}

\def\gcheck{\hbox{\lower2.5pt\hbox{\Bigcheck}\kern-8pt$\g$}}
\def\lhat{\hbox{\raise.5pt\hbox{\Bighat}\kern-8pt$\l$}}

\def\Fcheck{\kern2pt\hbox{\raise1pt\hbox{\Bigcheck}\kern-10pt{$\cal F$}}}
\def\Fhat{\kern2pt\hbox{\raise1pt\hbox{\Bighat}\kern-10pt{$\cal F$}}}

\def\boldIP{\relax{\tenboldrm \char '111 \kern-.18em \char '120}}
\def\boldcp#1{\relax{\boldIP\kern-2pt\lower.5ex\hbox{\eightboldmath #1}}}
\def\cp#1{\relax\ifmmode {\IP\kern-2pt{}_{#1}}\else $\IP\kern-2pt{}_{#1}$\fi}
\def\h#1#2{\relax\ifmmode {b_{#1#2}}\else $b_{#1#2}$\fi}
\def\Z{\hfill\break}
\def\imag{\Im m}
\def\half{{1\over 2}}

\def\frac#1#2{{#1\over #2}}

\def\pd#1#2{{\partial #1\over\partial #2}}

\def\cone{\relax\thinspace\hbox{$<\kern-.8em{)}$}}
\mathchardef\mho"0A30

\def\Psymbol#1#2{\hbox{\twelvecal\char'120}\left\{
                 \matrix{0&\infty&1\cr #1}\hskip8truept\matrix{#2}\right\}}

\def\asymp{\sim}
\def\-{\hphantom{-}}
\def\mirror{\ca{W}}


\def\npb#1{Nucl.\ Phys.\ {\bf B#1}}

\def\cmp#1{Commun. Math. Phys. {\bf #1}}
\def\plb#1{Phys. Lett. {\bf #1B}}


\def\picture #1 by #2 (#3){\vbox to #2{\hrule width #1 height 0pt depth 0pt
                                       \vfill\special{picture #3}}}
\def\scaledpicture #1 by #2 (#3 scaled #4){{\dimen0=#1 \dimen1=#2
           \divide\dimen0 by 1000 \multiply\dimen0 by #4
            \divide\dimen1 by 1000 \multiply\dimen1 by #4
            \picture \dimen0 by \dimen1 (#3 scaled #4)}}
\def\illustration #1 by #2 (#3){\vbox to #2{\hrule width #1 height 0pt depth
0pt
                                       \vfill\special{illustration #3}}}
\def\scaledillustration #1 by #2 (#3 scaled #4){{\dimen0=#1 \dimen1=#2
           \divide\dimen0 by 1000 \multiply\dimen0 by #4
            \divide\dimen1 by 1000 \multiply\dimen1 by #4
            \illustration \dimen0 by \dimen1 (#3 scaled #4)}}


\def\delaOssa{\nobreak\vskip1truein\hbox to\hsize
       {\hskip 4truein Xenia de la Ossa\hfill}}

\def\hoy{\number\day\space de \ifcase\month\or enero\or febrero\or marzo\or
       abril\or mayo\or junio\or julio\or agosto\or septiembre\or octubre\or
       noviembre\or diciembre\fi\space de \number\year}


\newif\ifproofmode
\proofmodefalse

\newif\ifforwardreference
\forwardreferencefalse

\newif\ifchapternumbers
\chapternumbersfalse

\newif\ifcontinuousnumbering
\continuousnumberingfalse

\newif\iffigurechapternumbers
\figurechapternumbersfalse

\newif\ifcontinuousfigurenumbering
\continuousfigurenumberingfalse

\newif\iftablechapternumbers
\tablechapternumbersfalse

\newif\ifcontinuoustablenumbering
\continuoustablenumberingfalse

\font\eqsixrm=cmr6

\def\marginstyle{\eqsixrm}

\newtoks\chapletter
\newcount\chapno
\newcount\eqlabelno
\newcount\figureno
\newcount\tableno

\chapno=0
\eqlabelno=0
\figureno=0
\tableno=0

\def\chapfolio{\ifnum\chapno>0 \the\chapno\else\the\chapletter\fi}

\def\bumpchapno{\ifnum\chapno>-1 \global\advance\chapno by 1
\else\global\advance\chapno by -1 \setletter\chapno\fi
\ifcontinuousnumbering\else\global\eqlabelno=0 \fi
\ifcontinuousfigurenumbering\else\global\figureno=0 \fi
\ifcontinuoustablenumbering\else\global\tableno=0 \fi}

\def\setletter#1{\ifcase-#1{}\or{}%
\or\global\chapletter={A}%
\or\global\chapletter={B}%
\or\global\chapletter={C}%
\or\global\chapletter={D}%
\or\global\chapletter={E}%
\or\global\chapletter={F}%
\or\global\chapletter={G}%
\or\global\chapletter={H}%
\or\global\chapletter={I}%
\or\global\chapletter={J}%
\or\global\chapletter={K}%
\or\global\chapletter={L}%
\or\global\chapletter={M}%
\or\global\chapletter={N}%
\or\global\chapletter={O}%
\or\global\chapletter={P}%
\or\global\chapletter={Q}%
\or\global\chapletter={R}%
\or\global\chapletter={S}%
\or\global\chapletter={T}%
\or\global\chapletter={U}%
\or\global\chapletter={V}%
\or\global\chapletter={W}%
\or\global\chapletter={X}%
\or\global\chapletter={Y}%
\or\global\chapletter={Z}\fi}

\def\tempsetletter#1{\ifcase-#1{}\or{}%
\or\global\chapletter={A}%
\or\global\chapletter={B}%
\or\global\chapletter={C}%
\or\global\chapletter={D}%
\or\global\chapletter={E}%
\or\global\chapletter={F}%
\or\global\chapletter={G}%
\or\global\chapletter={H}%
\or\global\chapletter={I}%
\or\global\chapletter={J}%
\or\global\chapletter={K}%
\or\global\chapletter={L}%
\or\global\chapletter={M}%
\or\global\chapletter={N}%
\or\global\chapletter={O}%
\or\global\chapletter={P}%
\or\global\chapletter={Q}%
\or\global\chapletter={R}%
\or\global\chapletter={S}%
\or\global\chapletter={T}%
\or\global\chapletter={U}%
\or\global\chapletter={V}%
\or\global\chapletter={W}%
\or\global\chapletter={X}%
\or\global\chapletter={Y}%
\or\global\chapletter={Z}\fi}

\def\chapshow#1{\ifnum#1>0 \relax#1%
\else{\tempsetletter{\number#1}\chapno=#1\chapfolio}\fi}

\def\chaplabel#1{\bumpchapno\ifproofmode\ifforwardreference
\immediate\write1{\noexpand\expandafter\noexpand\def
\noexpand\csname CHAPLABEL#1\endcsname{\the\chapno}}\fi\fi
\global\expandafter\edef\csname CHAPLABEL#1\endcsname
{\the\chapno}\ifproofmode\llap{\hbox{\marginstyle #1\ }}\fi\chapfolio}

\def\chapref#1{\ifundefined{CHAPLABEL#1}??\ifproofmode\ifforwardreference%
\else\write16{ ***Undefined Chapter Reference #1*** }\fi
\else\write16{ ***Undefined Chapter Reference #1*** }\fi
\else\edef\LABxx{\getlabel{CHAPLABEL#1}}\chapshow\LABxx\fi
\ifproofmode\write2{Chapter #1}\fi}

\def\eqnum{\global\advance\eqlabelno by 1
\eqno(\ifchapternumbers\chapfolio.\fi\the\eqlabelno)}

\def\eqlabel#1{\global\advance\eqlabelno by 1 \ifproofmode\ifforwardreference
\immediate\write1{\noexpand\expandafter\noexpand\def
\noexpand\csname EQLABEL#1\endcsname{\the\chapno.\the\eqlabelno?}}\fi\fi
\global\expandafter\edef\csname EQLABEL#1\endcsname
{\the\chapno.\the\eqlabelno?}\eqno(\ifchapternumbers\chapfolio.\fi
\the\eqlabelno)\ifproofmode\rlap{\hbox{\marginstyle #1}}\fi}

\def\eqalignnum{\global\advance\eqlabelno by 1
&(\ifchapternumbers\chapfolio.\fi\the\eqlabelno)}

\def\eqalignlabel#1{\global\advance\eqlabelno by 1 \ifproofmode
\ifforwardreference\immediate\write1{\noexpand\expandafter\noexpand\def
\noexpand\csname EQLABEL#1\endcsname{\the\chapno.\the\eqlabelno?}}\fi\fi
\global\expandafter\edef\csname EQLABEL#1\endcsname
{\the\chapno.\the\eqlabelno?}&(\ifchapternumbers\chapfolio.\fi
\the\eqlabelno)\ifproofmode\rlap{\hbox{\marginstyle #1}}\fi}

\def\eqref#1{\hbox{(\ifundefined{EQLABEL#1}***)\ifproofmode\ifforwardreference%
\else\write16{ ***Undefined Equation Reference #1*** }\fi
\else\write16{ ***Undefined Equation Reference #1*** }\fi
\else\edef\LABxx{\getlabel{EQLABEL#1}}%
\def\LAByy{\expandafter\stripchap\LABxx}\ifchapternumbers%
\chapshow{\LAByy}.\expandafter\stripeq\LABxx%
\else\ifnum\number\LAByy=\chapno\relax\expandafter\stripeq\LABxx%
\else\chapshow{\LAByy}.\expandafter\stripeq\LABxx\fi\fi)\fi}%
\ifproofmode\write2{Equation #1}\fi}

\def\fignum{\global\advance\figureno by 1
\relax\iffigurechapternumbers\chapfolio.\fi\the\figureno}

\def\figlabel#1{\global\advance\figureno by 1
\relax\ifproofmode\ifforwardreference
\immediate\write1{\noexpand\expandafter\noexpand\def
\noexpand\csname FIGLABEL#1\endcsname{\the\chapno.\the\figureno?}}\fi\fi
\global\expandafter\edef\csname FIGLABEL#1\endcsname
{\the\chapno.\the\figureno?}\iffigurechapternumbers\chapfolio.\fi
\ifproofmode\llap{\hbox{\marginstyle#1
\kern1.2truein}}\relax\fi\the\figureno}

\def\figref#1{\hbox{\ifundefined{FIGLABEL#1}!!!!\ifproofmode\ifforwardreference%
\else\write16{ ***Undefined Figure Reference #1*** }\fi
\else\write16{ ***Undefined Figure Reference #1*** }\fi
\else\edef\LABxx{\getlabel{FIGLABEL#1}}%
\def\LAByy{\expandafter\stripchap\LABxx}\iffigurechapternumbers%
\chapshow{\LAByy}.\expandafter\stripeq\LABxx%
\else\ifnum \number\LAByy=\chapno\relax\expandafter\stripeq\LABxx%
\else\chapshow{\LAByy}.\expandafter\stripeq\LABxx\fi\fi\fi}%
\ifproofmode\write2{Figure #1}\fi}

\def\tabnum{\global\advance\tableno by 1
\relax\iftablechapternumbers\chapfolio.\fi\the\tableno}

\def\tablabel#1{\global\advance\tableno by 1
\relax\ifproofmode\ifforwardreference
\immediate\write1{\noexpand\expandafter\noexpand\def
\noexpand\csname TABLABEL#1\endcsname{\the\chapno.\the\tableno?}}\fi\fi
\global\expandafter\edef\csname TABLABEL#1\endcsname
{\the\chapno.\the\tableno?}\iftablechapternumbers\chapfolio.\fi
\ifproofmode\llap{\hbox{\marginstyle#1
\kern1.2truein}}\relax\fi\the\tableno}

\def\tabref#1{\hbox{\ifundefined{TABLABEL#1}!!!!\ifproofmode\ifforwardreference%
\else\write16{ ***Undefined Table Reference #1*** }\fi
\else\write16{ ***Undefined Table Reference #1*** }\fi
\else\edef\LABtt{\getlabel{TABLABEL#1}}%
\def\LABTT{\expandafter\stripchap\LABtt}\iftablechapternumbers%
\chapshow{\LABTT}.\expandafter\stripeq\LABtt%
\else\ifnum\number\LABTT=\chapno\relax\expandafter\stripeq\LABtt%
\else\chapshow{\LABTT}.\expandafter\stripeq\LABtt\fi\fi\fi}%
\ifproofmode\write2{Table#1}\fi}

\newdimen\sectionskip     \sectionskip=20truept
\newcount\sectno
\def\section#1#2{\sectno=0 \null\vskip\sectionskip
    \centerline{\chaplabel{#1}.~~{\bf#2}}\nobreak\vskip.2truein
    \noindent\ignorespaces}

\def\advancesectno{\global\advance\sectno by 1}
\def\sectfolio{\number\sectno}
\def\subsection#1{\goodbreak\advancesectno\null\vskip10pt
                  \noindent\chapfolio.~\sectfolio.~{\bf #1}
                  \nobreak\vskip.05truein\noindent\ignorespaces}

\def\uttg#1{\null\vskip.1truein
    \ifproofmode \line{\hfill{\bf Draft}:
    UTTG--{#1}--\number\year}\line{\hfill\today}
    \else \line{\hfill UTTG--{#1}--\number\year}
    \line{\hfill\ifcase\month\or January\or February\or March\or April\or
May\or June
    \or July\or August\or September\or October\or November\or December\fi
    \space\number\year}\fi}

\def\getlabel#1{\csname#1\endcsname}
\def\ifundefined#1{\expandafter\ifx\csname#1\endcsname\relax}
\def\stripchap#1.#2?{#1}
\def\stripeq#1.#2?{#2}

%
\catcode`@=11 
\def\space@ver#1{\let\@sf=\empty\ifmmode#1\else\ifhmode%
\edef\@sf{\spacefactor=\the\spacefactor}\unskip${}#1$\relax\fi\fi}
\newcount\referencecount     \referencecount=0
\newif\ifreferenceopen       \newwrite\referencewrite
\newtoks\rw@toks
\def\refmark#1{\relax[#1]}
\def\refend{\refmark{\number\referencecount}}
\newcount\lastrefsbegincount \lastrefsbegincount=0
\def\refsend{\refmark{\count255=\referencecount%
\advance\count255 by -\lastrefsbegincount%
\ifcase\count255 \number\referencecount%
\or\number\lastrefsbegincount,\number\referencecount%
\else\number\lastrefsbegincount-\number\referencecount\fi}}
\def\refch@ck{\chardef\rw@write=\referencewrite
\ifreferenceopen\else\referenceopentrue
\immediate\openout\referencewrite=referenc.texauxil \fi}
%
{\catcode`\^^M=\active 
  \gdef\obeyendofline{\catcode`\^^M\active \let^^M\ }}%
%
{\catcode`\^^M=\active 
  \gdef\ignoreendofline{\catcode`\^^M=5}}
{\obeyendofline\gdef\rw@start#1{\def\t@st{#1}\ifx\t@st\blankend%
\endgroup\@sf\relax\else\ifx\t@st\bl@nkend\endgroup\@sf\relax%
\else\rw@begin#1
\backtotext
\fi\fi}}
{\obeyendofline\gdef\rw@begin#1
{\def\n@xt{#1}\rw@toks={#1}\relax%
\rw@next}}
\def\blankend{}
{\obeylines\gdef\bl@nkend{
}}
\newif\iffirstrefline  \firstreflinetrue
\def\rwr@teswitch{\ifx\n@xt\blankend\let\n@xt=\rw@begin%
\else\iffirstrefline\global\firstreflinefalse%
\immediate\write\rw@write{\noexpand\obeyendofline\the\rw@toks}%
\let\n@xt=\rw@begin%
\else\ifx\n@xt\rw@@d \def\n@xt{\immediate\write\rw@write{%
\noexpand\ignoreendofline}\endgroup\@sf}%
\else\immediate\write\rw@write{\the\rw@toks}%
\let\n@xt=\rw@begin\fi\fi\fi}
\def\rw@next{\rwr@teswitch\n@xt}
\def\rw@@d{\backtotext} \let\rw@end=\relax
\let\backtotext=\relax

\newdimen\refindent     \refindent=30pt
\def\Textindent#1{\noindent\llap{#1\enspace}\ignorespaces}
\def\refitem#1{\par\hangafter=0 \hangindent=\refindent\Textindent{#1}}
\def\REFNUM#1{\space@ver{}\refch@ck\firstreflinetrue%
\global\advance\referencecount by 1 \xdef#1{\the\referencecount}}
\def\refnum#1{\space@ver{}\refch@ck\firstreflinetrue%
\global\advance\referencecount by 1\xdef#1{\the\referencecount}\refend}

\def\REF#1{\REFNUM#1%
\immediate\write\referencewrite{%
\noexpand\refitem{#1.}}%
\begingroup\obeyendofline\rw@start}
\def\ref{\refnum\?%
\immediate\write\referencewrite{\noexpand\refitem{\?.}}%
\begingroup\obeyendofline\rw@start}
\def\Ref#1{\refnum#1%
\immediate\write\referencewrite{\noexpand\refitem{#1.}}%
\begingroup\obeyendofline\rw@start}
\def\REFS#1{\REFNUM#1\global\lastrefsbegincount=\referencecount%
\immediate\write\referencewrite{\noexpand\refitem{#1.}}%
\begingroup\obeyendofline\rw@start}

\def\REFSCON#1{\REF#1}

\def\cite#1{\refmark#1}
\catcode`@=12 
%

%

\font\elevenboldmath=cmmib10 at 11pt
\font\tenboldmath=cmmib10 at 10pt

\font\eightboldmath=cmmib10 at 8pt

\font\elevenboldrm=cmbx12 at 11pt
\font\tenboldrm=cmbx12 at 10pt
\font\twelvemath=cmmi12 at 12pt

\font\scriptss=cmss8 at 6pt

\def\deriv{{\rm d}}

\def\question{\leavevmode\raise2pt\hbox{?`}}
\def\cropen#1{\crcr\noalign{\vskip #1}}

\def\bar{\overline}

\def\txt{\textstyle}

%

\def\real{\hbox{$\Re e$}}

\def\F{\hbox{${}_2F_1$}}

\def\hatto{\hbox{$\longrightarrow$
    \kern-11pt\lower1.5pt\hbox{$\widehat{}$}\kern11pt}}

\def\tildeto{\hbox{$\longrightarrow$
    \kern-11pt\lower1.5pt\hbox{$\widetilde{}$}\kern11pt}}

\def\third#1{\raise1pt\hbox{${\scriptstyle{#1\over 3}}$}}

\def\1{{\bf 1}}
\def\2{{\bf 2}}
\def\3{{\bf 3}}
\def\4{{\bf 4}}
\def\rootmetric#1{\sqrt{g^{\scriptscriptstyle #1,\bar{#1}}} }

\def\Psymbol#1#2{\hbox{\twelvecal\char'120}\left\{
                 \matrix{0&\infty&1\cropen{4pt}
 #1}\hskip8truept\matrix{#2}\right\}}

\def\Z{\ca{Z}}
\def\Ztilde{\relax\ifmmode \tilde{\Z}\else $\tilde{\Z}$\fi}

\def\T{\ca{T}}
\proofmodefalse
%
%
%
\baselineskip=13pt
\parskip=2pt
%
%
\pageno=1
\chapternumberstrue
\figurechapternumberstrue
\tablechapternumberstrue
\ifproofmode
\immediate\openout2=allcrossreferfile \fi
\ifforwardreference\input labelfile
\ifproofmode\immediate\openout1=labelfile \fi\fi
\noblackboxes
\hfuzz=1pt
\vfuzz=2pt
%
\expandafter \def \csname CHAPLABELZintro\endcsname {1}
\expandafter \def \csname EQLABELMW\endcsname {1.1?}
\expandafter \def \csname CHAPLABELZmanifolds\endcsname {2}
\expandafter \def \csname EQLABELscaling\endcsname {2.1?}
\expandafter \def \csname EQLABELres\endcsname {2.2?}
\expandafter \def \csname EQLABELeval\endcsname {2.3?}
\expandafter \def \csname EQLABELmonos\endcsname {2.4?}
\expandafter \def \csname TABLABELmonomials.Z\endcsname {2.1?}
\expandafter \def \csname CHAPLABELZYukawas\endcsname {3}
\expandafter \def \csname EQLABELtop\endcsname {3.1?}
\expandafter \def \csname EQLABELideal\endcsname {3.2?}
\expandafter \def \csname EQLABELdeal\endcsname {3.3?}
\expandafter \def \csname EQLABELfcoup\endcsname {3.4?}
\expandafter \def \csname TABLABELconditions.Z\endcsname {3.1?}
\expandafter \def \csname EQLABELJofphi\endcsname {3.5?}
\expandafter \def \csname EQLABELesquare\endcsname {3.6?}
\expandafter \def \csname CHAPLABELZmodular\endcsname {4}
\expandafter \def \csname EQLABELlambda\endcsname {4.1?}
\expandafter \def \csname EQLABELZJofphi\endcsname {4.2?}
\expandafter \def \csname EQLABELZchiexpand\endcsname {4.3?}
\expandafter \def \csname EQLABELZ1/tau\endcsname {4.4?}
\expandafter \def \csname EQLABELZphioftau\endcsname {4.5?}
\expandafter \def \csname EQLABELZtauofphi\endcsname {4.6?}
\expandafter \def \csname EQLABELZwtransf\endcsname {4.7?}
\expandafter \def \csname EQLABELZsumtransf\endcsname {4.8?}
\expandafter \def \csname TABLABELtransf.Z\endcsname {4.1?}
\expandafter \def \csname EQLABELZsfn\endcsname {4.9?}
\expandafter \def \csname EQLABELZmufn\endcsname {4.10?}
\expandafter \def \csname CHAPLABELZperiods\endcsname {5}
\expandafter \def \csname EQLABELZintersect.nos.\endcsname {5.1?}
\expandafter \def \csname EQLABELperiodint\endcsname {5.2?}
\expandafter \def \csname EQLABELmutbar\endcsname {5.3?}
\expandafter \def \csname EQLABELomegagauge\endcsname {5.4?}
\expandafter \def \csname EQLABELZp\endcsname {5.5?}
\expandafter \def \csname EQLABELtorusp\endcsname {5.6?}
\expandafter \def \csname EQLABELZhge\endcsname {5.7?}
\expandafter \def \csname EQLABELZs2torus\endcsname {5.8?}
\expandafter \def \csname EQLABELZs1torus\endcsname {5.9?}
\expandafter \def \csname EQLABELPieight\endcsname {5.10?}
\expandafter \def \csname EQLABELZdiffq\endcsname {5.11?}
\expandafter \def \csname EQLABELPFtwo\endcsname {5.12?}
\expandafter \def \csname EQLABELQR\endcsname {5.13?}
\expandafter \def \csname EQLABELAandC\endcsname {5.14?}
\expandafter \def \csname EQLABELPiten\endcsname {5.15?}
\expandafter \def \csname EQLABELZfchoice\endcsname {5.16?}
\expandafter \def \csname CHAPLABELZYukawanorm\endcsname {6}
\expandafter \def \csname TABLABELconditions.cf\endcsname {6.1?}
\expandafter \def \csname CHAPLABELappendix1\endcsname {-2}
\expandafter \def \csname EQLABELZhgfrelations\endcsname {-2.1?}
\expandafter \def \csname EQLABELZperiodones\endcsname {-2.2?}
\expandafter \def \csname EQLABELZhgfd\endcsname {-2.3?}
\nopagenumbers
\null\vskip-.75truein
\rightline{CERN-TH.6831/93}
\rightline{UTTG-24-92}
\rightline{March 1993}
\vskip .5truein
\centerline{\seventeenrm GENERALIZED CALABI-YAU MANIFOLDS
            \vphantom{\raise1ex\hbox{*}}}
\vskip .1truein
\centerline{\seventeenrm AND THE MIRROR OF A RIGID MANIFOLD
\footnote{\raise1ex\hbox{*}}{\eightrm Supported in part
       by the Robert A. Welch Foundation and N.S.F. Grant
       PHY9009850.
       \medskip
       \leftline{CERN-TH.6831/93}\vskip-3pt\leftline{UTTG-24-92}}}
\vskip .4truein
\centerline {P. Candelas $^{1,3}$,~E. Derrick $^{2,3}$~and~L. Parkes $^3$ }
\vskip .4truein
\line{\hfil
\vtop{\hsize = 2.0truein
\centerline {\it $^1$Theory Division}
\centerline {\it  CERN}
\centerline {\it CH 1211 Geneva 23}
\centerline {\it Switzerland} }
\hskip5pt
\vtop{\hsize = 2.0truein
\centerline {\it $^2$ Institut de Physique}
\centerline {\it  Universit\'e de Neuch\^atel}
\centerline {\it CH 2000 Neuch\^atel}
\centerline {\it Switzerland} }
\hskip15pt
\vtop{\hsize = 2.0truein
\centerline {\it $^3$ Theory Group}
\centerline {\it Department of Physics}
\centerline {\it The University of Texas}
\centerline {\it Austin, TX 78712, U.S.A.}}
\hfil}
\vskip .6truein
\bigskip
\centerline{\bf ABSTRACT}
\vskip.2truein{\vbox{\baselineskip = 12pt
\noindent The \Z\ manifold is a \cym\ with $b_{21}=0$.  At first sight it
seems to provide a counter example to the mirror hypothesis since its mirror
would have $b_{11}=0$ and hence could not be \K.  However by identifying
the \Z\ manifold with the Gepner model $1^9$ we are able to ascribe a
geometrical interpretation to the mirror, \Ztilde, as a certain seven-
dimensional manifold.  The mirror manifold \Ztilde\ is a representative
of a class of generalized \cys, which we describe, that can be realized
as manifolds of dimension five and seven.  Despite their dimension
these generalized \cys\ correspond to superconformal theories with
$c=9$ and so are perfectly good for compactifying the heterotic
string to the four dimensions of space-time.  As a check of mirror symmetry
we compute the structure of the space of complex structures of
the mirror \Ztilde\ and check that this reproduces the known results
for the Yukawa couplings and metric appropriate to the \K\ class parameters
on the \Z\ orbifold together with their instanton corrections.
In addition to reproducing known results we can calculate the periods of
the manifold to arbitrary order in the blowing up parameters.
This provides a means of calculating the Yukawa couplings and metric
as functions also to arbitrary order in the blowing up parameters which is
difficult to do by traditional~methods. }}
\nopagenumbers
\newpage

\section{Zintro}{Introduction}
\pageno=1
\footline={\hss\tenrm\folio\hss}
\noindent
The existence of mirror symmetry among \cys\ is the fact that,
roughly speaking, \cys\ come in so-called mirror pairs.
The mirror operation can be thought of as a reflection of the
Hodge diamond for a manifold \M
$$
\llap{$b_{pq} ~=~~$}
\vcenter{\settabs\+\h 11&\h 11&\h 11&\h 11&\h 11&\h 11&\cr
                      \+~~&~~&~~&~1&~~&~~&~~\cr
                      \+~~&~~&~0&~~&~0&~~&~~\cr
                      \+~~&~0&~~&\h 11&~~&~0&~~\cr
           \+~1$\!$&~~&\h 21&~~&\h 21&~~&~$\!$1~~\cr
                      \+~~&~0&~~&\h 11&~~&~0&~~\cr
                      \+~~&~~&~0&~~&~0&~~&~~\cr
                      \+~~&~~&~~&~1&~~&~~&~~\cr}
                      $$
about a diagonal axis.  The effect is to exchange the values of
$b_{11}$ and $b_{21}$ so that the corresponding Hodge numbers
for the mirror \mirror\ are given by
$$ \h 11(\mirror) = \h 21(\M) ~~,~~
   \h 21(\mirror) = \h 11(\M) ~. \eqlabel{MW} $$
A better statement of mirror symmetry is that \cys\ are
realizations of N=2 superconformal field theories
and that a given SCFT can be realized as a \cym\
in two different ways \M, \mirror,
whose Hodge numbers are related by \eqref{MW}.
In the underlying SCFT there is no natural way to
decide which operators correspond to (1,1)-forms
and which to (2,1)-forms in an associated \cym.
It was this that led to the conjecture of the existence
of mirror symmetry whereby each SCFT would correspond to
a pair of \cys\ in which the roles of these two
types of forms would be exchanged \
\REFS{\Dixon}{L.~Dixon, in {\sl Superstrings,
Unified Theories, and Cosmology 1987}, (G.~Furlan
{\it et. al.}, eds.) World Scientific, 1988, p. 67.}
\REFSCON{\LVW}{W.~Lerche, C.~Vafa and N.~Warner,
\npb{324} (1989) 427.}\refsend.

It is an essential fact that \cys\ have parameters
corresponding to the possible deformations of the complex
structure and of the \K\ class.
Infinitesimal deformations of the complex structure
of \M\ are in one-one correspondence with the
elements of the cohomology group
\H 21(\M) while the infinitesimal deformations of the
\K\ class are in one-one correspondence with the
elements of \H 11(\M).
Under mirror symmetry, these two parameter spaces are
exchanged.
The realization that both types of parameter space
have the same structure, both being described by
special geometry \
\REFS{\AS}{A.~Strominger, \cmp{133} (1990) 163.}
\REFSCON{\Cd}{P.~Candelas and X.~C.~de~la~Ossa,
\npb{355} (1991) 455.}\refsend,
lent strong evidence to the mirror symmetry hypothesis.
A construction of a large class of \cys\ revealed that
the great bulk of the manifolds so constructed occur in
mirror pairs~
\Ref{\CLS}{P.~Candelas, M.~Lynker and R.~Schimmrigk, \npb {341}
(1990) 383.}.
Contemporaneous with this was the construction of Greene and Plesser \
\Ref{\GP}{B.~R.~Greene and M.~R.~Plesser, \npb{338} (1990) 15.}\
who, by means of exploiting the correspondence between
certain manifolds and the Gepner models,
were able to provide a construction of the mirrors for these
cases.
A subsequent calculation by Aspinwall, L\"utken and Ross \
\Ref{\ALR}{P.~Aspinwall, A.~L\"utken and G.~G.~Ross, \plb {241} (1990) 373.}\
identified a large complex structure limit in which,
for a certain mirror pair,
the Yukawa coupling appropriate to the complex structures
of the mirror goes over to the topological value of the
Yukawa coupling appropriate to the \K\ class of the original manifold.

The precise circumstances under which mirror symmetry is true
are not known and there is currently no generally applicable procedure
for constructing the mirror of a given manifold,
though a number of procedures are applicable to special
classes of manifolds \
\REF{\BH}{ P.~Berglund and T.~H\"ubsch, {\sl A Generalized
Construction of Mirror Manifolds}, UTTG--32--91,
HUTMP--91/B319, CERN--TH--6341/91.}
\cite{{\CLS,\GP,\ALR,\BH}}.
There is also, at first sight, an immediate class of
counter examples furnished by the rigid manifolds.
These are manifolds for which $\h 21(\M)=0$.
The mirror would seemingly have to have $\h 11(\mirror)=0$ and
hence could not be \K.  The prototypical example of a rigid
\cym\ is the so-called \Z\ manifold \
\Ref{\CHSW}{P.~Candelas, G.~Horowitz, A.~Strominger and E.~Witten,
\npb{258} (1985) 46.}\
for which $\h 21=0$ and $\h 11 = 36$.
It turns out, however, that we can construct a mirror for the \Z\
as a {\bf seven}--dimensional manifold with positive first Chern class.
We will see that the parameter space of complex structures of
the mirror \Ztilde\ is described by special geometry
and that we can obtain the quantum corrections to the
Yukawa couplings and the kinetic terms of the low energy theory
that results from the compactification of string theory on the
\Z\ manifold by studying the space of complex structures of
\Ztilde.
The \Z\ manifold has an orbifold limit
which has been studied extensively and the instanton
contributions to the Yukawa couplings have already been
computed \
\REFS{\HV}{S.~Hamidi and C.~Vafa, \npb{279} (1987) 465.}
\REFSCON{\DFMS}{L.~Dixon, D.~Friedan, E.~Martinec, and S. ~Shenker,
\npb{282} (1987) 13.}
\REFSCON{\LMN}{J.~Lauer, J.~Mas and H.~P.~Nilles, \plb{226} (1989) 251;
\npb{351} (1991) 353.}
\refsend.
For the orbifold limit we derive nothing new except that
we check that we recover the standard results by means of mirror
symmetry.  We can also go further than has hithero proved
possible with traditional methods since mirror symmetry allows
us to find the Yukawa couplings, say,
even away from the orbifold limit.  We do not attempt
a full treatment involving all 27 parameters associated
with the resolution of the singularities;
rather, we allow just one of these 27 parameters to
vary away from the orbifold limit.
It is clear that more complicated cases involving more
parameters are amenable to study
however our main interest here is to check that mirror symmetry
is applicable and gives correct answers.

The paper is organized as  follows:
in Section~\chapref{Zmanifolds} we describe the \Z\ manifold,
its mirror \Ztilde\ and the class of generalized \cys\ to which it belongs.
In Section~\chapref{ZYukawas} we calculate the Yukawa couplings
of \Ztilde\ by means of a calculation in the ring of the defining polynomial
following a method described in \
\Ref{\PC}{P.~Candelas, \npb {298} (1988) 458.}.
The issue of the normalization of the couplings
is addressed via the techniques of special geometry.
For this we need to find an integral basis for the periods on the manifold.
In Section~\chapref{Zperiods} we calculate
periods both at and away from the orbifold limit.
In the orbifold limit the close relation between \Ztilde\
and the product of tori can be used to
find the integral basis.
Away from this limit we must check that the modular group
acts on the period vector by integral matrices. This requirement however
does not fix the basis uniquely and we are obliged to describe explicitly
a homology basis. Being somewhat technical this part of the analysis is
relegated to an appendix. We investigate the actions of the modular group in
Section~\chapref{Zmodular}, and we settle
on a proper basis for the periods in Section~\chapref{Zperiods}.5.
Finally in Section~\chapref{ZYukawanorm}
we calculate the normalized Yukawa couplings on \Ztilde\
and compare to previous calculations. The reason that we make the comparison
with the orbifold limit only after this
lengthy discussion of the properties of\vadjust{\vfil\eject}
the manifold away from the orbifold limit is that the most involved part
of the calculation is the computation of the normalization factors.
For example, if $s$ denotes a parameter associated with blowing up
the orbifold singularities then the proper normalization of the
Yukawa coupling $y_{sss}$, say, involves a knowledge of the metric
component $g_{s\bar{s}}$ and this in turn requires the computation of
the \K\ potential and hence the periods as functions~of~$s$.
%
\section{Zmanifolds}{The \Z\ Manifold, its Mirror and a Class of
Generalized Calabi-Yau Manifolds}
\noindent
Recall the construction of the \Z\ manifold
\REF{\SW}{A. Strominger and E. Witten, \cmp{101} (1985) 341.}
\ \cite{{\CHSW,\SW}}.
Let \T\ be the product of 3 tori
$$
\T=\T_1\times \T_2\times \T_3~,$$
with each torus formed by making the identifications
$$
z_i \simeq z_i + 1 \simeq z_i + \o^{1/2}~~, ~~\o = e^{2 \p i/3}~.$$
The Euler number of \T\ is zero.
If we divide \T\ by the $\IZ_3$ generated by
$z_i \rightarrow \o z_i$,
we obtain the \Z\ orbifold.
Each torus has the three fixed points
\hbox{$r \sqrt{\third 1} \o^{1/4}$}, $r=0,1,2$, and
\T\ has the 27 fixed points
$$ f_{mnp} = (m,n,p) {\o^{1/4} \over \sqrt3}~\quad,\quad~m,n,p=0,1,2~.$$
We delete the fixed points and glue in
appropriate Eguchi-Hansen balls
(which have $\ch = 3$) to obtain the \Z\ manifold.
The Euler number of the \Z\ is therefore
$$
\ch(\Z) = {0-27 \over 3} + 27 \cdot 3 = 72~.$$
It has Hodge numbers $b_{11}(\Z) = 36$ and $b_{21}(\Z)=0$.
The counting is that 9 of the (1,1)--forms
can be thought of as the forms
$e_{i \bar \jmath} = dz^i \wedge dz^{\bar \jmath}$
that descend from \T.
The other 27 (1,1)--forms come from blowing
up the fixed points $ f_{mnp} $.

The \Z\ manifold corresponds to the $1^9$ Gepner model \
\REFS{\Gep}{D.~Gepner, \npb{296} (1987) 757.}
\REFSCON{\LR}{C.~A.~L\"utken and G.~G.~Ross, \plb{214} (1988) 357.}
\REFSCON{\GVW}{B.~R.~Greene, C.~Vafa and N.~Warner,
\npb{324} (1989) 371.}
\refsend \
and corresponds to a Landau-Ginzburg potential
$$ p=\sum_{k=1}^9 y_k^3~.$$
It is natural to think of the
nine $y_k$ as the homogeneous coordinates of \cp8
and therefore to think of the mirror manifold \Ztilde\
as \cp8[3], a hypersurface of degree 3 in \cp8.
This is a seven--dimensional manifold.
It is of interest to compute the Hodge decomposition of
the middle cohomology, $H^7$.
In the following we shall be concerned with the quotient
of \cp8[3]\  by various groups $G$.
The Hodge decomposition of $H^7$ for \cp8[3]/$G$ is
$$
\llap{$H^7~:$}
{}~ 0 ~~0~~\!\!\lower2pt\hbox{$\boxed{1~~$\b$~~$\b$~~1}$}~~0~~0~~.  $$
The dimension of $H^{5,2}$ is always 1 while the dimension of
$H^{4,3}$ assumes various values $\b$ depending on $G$.
For the case of \cp8[3]\ itself $\b=84$.
The important point is that the boxed entries have the same form
as the Hodge decomposition of $H^3$ for a \cym.
In particular, \cp8[3] has a unique (5,2)--form $\O_{5,2}$
which is the analogue of the holomorphic three--form $\O_{3,0}$
familiar from the theory of \cys.
In the study of the parameter space of the complex structures
of a \cym\ interest focusses on the variation of the periods of
the holomorphic three--form \
\cite{{\AS,\Cd}}.
In this way one learns that the space of complex structures is
described by special geometry.  The same is true for the
space of complex structures of our sevenfold \cp8[3]
owing to the fact that there is a unique (5,2)--form
and the (4,3)--forms correspond to the variation of $\O_{5,2}$
with respect to the complex structure.
To understand the existence of $\O_{5,2}$ recall the following
construction \cite\CLS\ of the holomorphic three--form for a \cym\
that is presented as $\IP{}_4^{(k_1,\ldots,k_5)}[d]$.
This is the vanishing locus of a polynomial $p$ in a weighted projective
space \cp4 with coordinates $(y_1,\ldots,y_5)$
that have weights $(k_1,\ldots,k_5)$ {\it i.e.} the coordinates
are identified
$$
  (y_1,y_2,\ldots,y_5) \simeq (\l^{k_1} y_1, \l^{k_2} y_2,
          \ldots, \l^{k_5} y_5) \eqlabel{scaling}$$
for any $\l \neq 0$.
We set
$$
\m = {1\over 4!}
\e_{A_1 A_2 A_3 A_4 A_5} y^{A_1} dy^{A_2} dy^{A_3} dy^{A_4} dy^{A_5}$$
and construct $\O_{3,0}$ as a residue by dividing $\m$ by $p$
and taking an integral around a one-dimensional contour $C_p$
that winds around the hypersurface $p=0$ in the embedding \cp4
$$
\O_{3,0}={1\over{2 \p i}} \int_{C_p} {\m\over p}~.$$
This construction makes sense because $\m/p$ is invariant
under the scaling \eqref{scaling}.
Invariance under scaling requires the degree of $p$ to be related
to the weights by
$$  \sum_{j=1}^5 k_j = d $$
but this is precisely the condition of vanishing first Chern class.

For the case of \cp8[3] we take
$$
\m ={1\over8!}\e_{A_1 A_2 ...A_9} y^{A_1} dy^{A_2} \ldots dy^{A_9}$$
so $\m$ scales with $\l^9$.  Since the cubic $p$ scales with $\l^3$
we must divide $\m$ by $p^3$ to obtain a form invariant
under the scaling.  Thus we construct
$$
\O_{5,2}={1\over{2 \p i}} \int_{C_p} {\m\over p^3} ~. \eqlabel{res}$$

It remains to explain why $\O_{5,2}$ is a (5,2)--form.
This is due to the fact that we have a third order pole.
A first order pole would have produced a (7,0)--form
and a second order pole would have produced a (6,1)--form.
A more correct statement is that the residue formula \eqref{res}
is true in cohomology and that the (7,0) and (6,1) parts
of the residue are exact.

The chiral ring can now be built up as in the previous case.
If $q$, $r$, and $s$ are three cubics then
${1\over{2 \p i}} \int {\m q \over p^4}$
is a (4,3)--form,
${1\over{2 \p i}} \int {\m q r \over p^5}$
is a (3,4)--form and
${1\over{2 \p i}} \int {\m q r s\over p^6}$
is a multiple of the unique (2,5)--form.
If we write $\left< \O,\bar\O \right>$ for $-i \int\O\wedge\bar\O$
then we have the familiar expression for the
Yukawa coupling, $\k(q,r,s)$, of the polynomials
$$
  \int_{C_p}  {\m q r s\over p^6}\,\bigg|_{2,5} = \k(q,r,s) \
         {\bar\O \over \left< \O,\bar\O \right>} ~.$$

The point that is being made is a general one not
restricted to the mirror of the \Z\ manifold.
Consider $\IP{}_{2m+2}^{(k_1,\ldots,k_{2m+3})}[d]$,
a hypersurface in a weighted projective space of
dimension $2m+2$ defined by a polynomial $p$ of degree $d$
with the degree related to the weights by
$$    \sum_{j=1}^{2m+3} k_j = md ~. \eqlabel{eval} $$
The middle dimensional cohomology of this manifold has the
structure
$$\eqalign{
\llap{$H^{2m+1}~:~$}
&\underbrace{0 \cdots 0}  ~~
\!\lower2pt\hbox{$\boxed{1~~$\b$~~$\b$~~1}$}
{}~~ \underbrace{0 \cdots 0} \cr
& \,m-1 \hphantom{  ~~
\!\lower3pt\hbox{$\boxed{1~~$\b$~~$\b$~~1}$}
{}~~ } \,m-1 \cr}
 $$
so we again find the special geometry structure sitting within it.
The reason is as before:  we have a unique
$(m+2,m-1)$--form given by the residue formula
$\O = {1 \over 2 \pi i} \int {\m \over p^m}$.
In the language of SCFT, Equation \eqref{eval} is the
statement that these theories have $c=9$ and so are
consistent string compactifications
(even though the dimension of the manifold is $2m+1$
which is not 3 unless we are dealing with the traditional
case of $m=1$).
There are many such manifolds. In \
\REFS{\KS}{A.~Klemm and R.~Schimmrigk,
{\sl Landau-Ginsburg Vacua}, CERN-TH-6459/92, HD-THEP-92-13.\newpage}
\REFSCON{\KrSk}{M.~Kreuzer and H.~Skarke,
{\sl No Mirror Symmetry in Landau Ginsburg Vacua},
CERN-TH-6461/92.}
\refsend\
the authors compile an exhaustive list of 3,284
such spaces for the cases $m=2,3$ and observe that higher values of $m$ yield
essentially nothing new. The reason is the fact that the extra variables
enter the defining polynomial only as quadratic terms and these are trivial
if the polynomial is thought of as a Landau--Ginzburg potential. A caveat
to this is that the extra quadratic terms can introduce $\IZ_2$ torsion
\ref{M.~Kreuzer, private communication.}.
It is an interesting and important question
whether it is possible to associate a three--dimensional manifold with each
member of this class. For recent results along these lines see \
\REFS{\Berglund}{P.~Berglund, {\sl Dimensionally Reduced Landau--Ginzburg
Orbifolds with Discrete Torsion}, NSF--ITP--93--27, UTTG--07--93.}
\REFSCON{\Schimm}{R.~Schimmrigk,
{\sl Critical Superstring Vacua from Noncritical Manifolds}, HD-THEP-82-92.}
\refsend.
We have come across this class of vacua
by considering the mirror of the \Z\ manifold.
Only a few members of this class are the mirrors
of rigid manifolds but it is not surprising to find
that the mirror of a rigid manifold is one of the generalized
\cys\ since it could not, after all, be a traditional \cym.
The notion that this class of manifold is important to the consideration
of the mirrors of rigid \cys\ was known independently to Vafa \
\Ref{\Vafa}{C.~Vafa,  in {\sl Essays on Mirror Manifolds}, (S.-T.~Yau ed.)
International Press, 1992, p. 96.}.

In virtue of the identification of the \Z\ manifold with the
SCFT $1^9$, together with the work of \
\Ref{\AL}{P.~Aspinwall and A.~L\"utken, \npb {353} (1991) 427.\hfill\break
P.~Aspinwall and A.~L\"utken, \npb {355} (1991) 482.} ,
we know that the mirror of the \Z\ is the quotient of \cp8[3] by
a group $G$ which is the $\IZ_3$ generated by $\z=(1 1 1~2 2 2~0 0 0)$,
where the notation means that
$$\z~:~y_i \rightarrow
\left\{ \eqalign{\o \, y_i ~~&,~~1\leq i\leq 3 \cr
               \o^2 y_i ~~&,~~4\leq i\leq 6 \cr
                y_i ~~&,~~7\leq i\leq 9 \cr}  \right.  $$
(Quotients of $1^9$ and their relation to the \Z\ were also discussed in
\ \Ref{\FIQ}{A.~Font, L.~E.~Ib\'a\~nez and F.~Quevedo,
\plb{224} (1989) 79,\hfill\break
E.~J.~Chun, J.~Lauer and H.~P.~Nilles,
Int.~J.~Mod.~Phys.~{\bf A7} (1992) 2175.}.)
In this case, the seventh cohomology decomposes in the same way as for
\cp8[3], but with $\b = 36$;  30 elements are polynomial
deformations while the other 6 come from smoothing the
quotient singularities.
To find the 30 elements which correspond to polynomial deformations we
recall that two deformations are the same if they differ by an
element of the ideal generated by the partial derivatives of the
defining polynomial, $\partial p \over \partial y_k$ \ \cite\PC.
In our case the generators of the
elements in the ideal are $y_k^2$, hence the deformations
must be of the form $y_i y_j y_k$ with $i, j,$ and
$k$ all different.  There are
${9!\over{6! 3!}} = 84$ such monomials.
In order to discuss the action of the $\IZ_3$ generator $\z$
it is convenient to think of
the 9 coordinates as the elements of a
3 by 3 matrix, so we set $x_{ij} = y_{3i+j-3}$,
where $i$ and $j$ now take the
values 1, 2 and 3.  Now we have
$$
p= \sum_{i,j=1}^3 x_{ij}^3 ~.$$
The monomial deformations of $p$
are displayed in Table~\tabref{monomials.Z} which also indicates which
are invariant
under the action of the $\IZ_3$ generator
$\z\,:\,x_{ij} \rightarrow \o^i x_{ij}$.
Note that there are indeed 84 total deformations before dividing by
$G$ and that 30 of them are invariant under $G$.

We can now identify these 30 monomials
invariant under $G$ with (1,1)--forms on the \Z\ manifold.
$$
e_i \simeq x_{i1}\, x_{i2}\, x_{i3}~,~~~
f_{mnp} \simeq x_{1m}\, x_{2 n}\, x_{3 p}~.\eqlabel{monos}$$
The off--diagonal $e_{i \bar \jmath}$'s from the \Z\ manifold
correspond to blow--ups on the mirror manifold,
and cannot be represented by polynomial deformations.
With this caveat, we write the polynomial
defining the mirror, \Ztilde, as
$$
p= \sum _{i,j} x_{ij}^3 - 3 \sum _k \ph_k e_k -
                          3 \sum _{m,n,p}s_{mnp}f_{mnp}~.$$
It would be of considerable interest to rectify this situation by finding
a way to represent the `missing' parameters. For our immediate purposes
however this limitation is not serious since we can discuss the theory with
$(b_{11},b_{21}) = (84,0)$ for which all the parameters may be reresented by
polynomial deformation or we may discuss the $\ca{Z}/\IZ_3$ manifold which
has $(b_{11},b_{21}) = (12,0)$ for which the same is true.
\vskip20pt
\vbox{\def\midrule{\vrule height 20pt depth 20pt}
$$
\vbox{\offinterlineskip\halign{\strut
#&$#$\quad\hfil
&\vrule\hfil\quad$#$\quad\hfil
&\vrule\hfil\quad$#$\quad\hfil\vrule\cr
\noalign{\hrule}
&\vrule height 15pt depth 10pt\hfil\quad \rm{monomial} &\rm{total\ number}
    &\rm{invariant\ under\ }$$G$$\cr
\noalign{\hrule\vskip3pt\hrule}
&\midrule \hfil\quad$$x_{i1}\,x_{i2}\,x_{i3}$$ &  3   &  3 \cr
\noalign{\hrule}
&\midrule \hfil\quad$$x_{im}\,x_{in}\,x_{jp}$$ &  54  &  0 \cr
\noalign{\hrule}
&\midrule \hfil\quad$$x_{1m}\,x_{2n}\,x_{3p}$$ &  27  & 27 \cr
\noalign{\hrule}}}$$
\tablecaption{monomials.Z}{The enumeration of the polynomial
deformations for the mirror of the \Z\ manifold.} }
%
\section{ZYukawas}{The Yukawa Couplings}
\noindent
We would like to calculate the Yukawa couplings
as a function of the $\ph_i$'s and $s_{mnp}$'s.
Initially let us simplify the problem by taking all $s_{mnp}= 0.$
The calculation is more complicated when we consider
$s_{mnp}\neq 0$, though there is conceptually no difference.

The classical or topological values of the couplings are given by the
intersection cubic on $H^2(\Z)$ \ \cite\SW\
$$ y_0 (a,b,c) = \int a\wedge b\wedge c ~. $$
$y_0$ counts the number of points of intersection of the three
four-surfaces that are dual to the two-forms $a$, $b$ and $c$.
In the orbifold limit the two-forms are the
$e_{i \bar \jmath} = dz^i \wedge dz^{\bar \jmath}$
and the 27 two-forms associated with the resolution of the fixed
points $f_{mnp}$.  In this limit these forms are supported on
the fixed points and we will, with a slight abuse of notation,
denote these forms by $f_{mnp}$ also.
We denote by $e_i$ the three diagonal forms
$dz^i \wedge dz^{\bar \imath}$
which have a direct correspondence with monomials in the mirror,
and by considering the intersections of the associated hypersurfaces
we immediately see that
$$\eqalign{
e_1 e_2 e_3 &\neq 0 \cr
e_i^2 &=0 \cr
e_i f_{mnp} &=0 \cr
f_{m_1 n_1 p_1} f_{m_2 n_2 p_2} f_{m_3 n_3 p_3} &=0
{}~~{\rm unless}
{}~~(m_1,n_1,p_1)=(m_2,n_2,p_2)=(m_3,n_3,p_3) ~.\cr} \eqlabel{top}$$

It is instructive to compare the topological couplings with the
same couplings derived from the mirror.
These couplings will contain all the sigma model corrections
to \eqref{top}.
We now think of the $e_i$ and the $f_{mnp}$ as the monomials
\eqref{monos} and we are to calculate the products modulo the
ideal generated by $\pd p {x_{ij}} =0$, \ie by the equations
$$ x_{ij}^2 = \ph_i \, x_{i,j+1} x_{i,j+2} ~. \eqlabel{ideal}$$
Let us first calculate $e_i^2$.  Suppressing the $i$ index we have
$$ e = x_1 x_2 x_3 $$
and the generators of the ideal are the equations
$$ x_j^2 = \ph \, x_{j+1} x_{j+2} ~. \eqlabel{deal} $$
We find
$$\eqalign{
e^2 &= x_1^2 x_2^2 x_3^2 \cr
    &= (\ph\,x_2 x_3) (\ph\,x_3 x_1)(\ph\,x_1 x_2) \cr
    &= \ph^3\,e^2 ~.\cr} $$
Thus $e^2$ vanishes since $\ph^3\neq 1$ in general.
In a similar way we see that $e_i f_{mnp}$ also vanishes.
Thus we have shown that the second and third of Equations \eqref{top}
are not corrected by instantons.

Consider now the $f^3$ couplings and label the indices on the three $f$'s such
that $m_{jk}$ is the $k$'th index on the $j$'th $f$, \ie the quantity of
interest is
$$
f_{m_{11}m_{12}m_{13}}f_{m_{21}m_{22}m_{23}}f_{m_{31}m_{32}m_{33}}~.
\eqlabel{fcoup}$$
To calculate this coupling we first gather together the factors
$x_{1 m_{11}} x_{1 m_{21}} x_{1 m_{31}}$ associated with the first index
on each $f$.  Again supressing the first index on each $x$
we are to calculate a product of the form $x_j x_k x_l$.
There are three possibilities:
$$\eqalign{
& x_1 x_2 x_3 \cr
& x_j^2 x_k = 0 ~~,~~~(j\neq k) \cr
& x_j^3 = \ph\,x_1 x_2 x_3 ~,\cr}$$
the right-hand sides of these relations following by virtue of \eqref{deal}.
Proceeding in this way we find that the cubic $f$ coupling \eqref{fcoup} is
proportional to $e_1e_2e_3$ with a constant of proportionality given by
the product of three factors, one corresponding to each triple of corresponding
indices, that is to each column of the matrix $m_{jk}$. These are given in
Table~\tabref{conditions.Z} and agree with the results in \
\cite\LMN\ and \
\Ref{\CMLN}{E.~J.~Chun, J.~Mas, J.~Lauer and H.~P.~Nilles, \plb{233}
(1989) 141.},
which are calculated by means of an instanton sum in the
SCFT of the orbifold.

\vbox{
\vskip 10pt
$$
\vbox{\offinterlineskip\halign{
&\vrule#&\strut \quad #\quad\hfil
&\vrule#&\quad\hfil #\hfil\quad&\vrule#\cr
\noalign{\hrule}
height 2pt&\omit&&\omit&\cr
height 12pt depth 5pt&\hfil Condition&&Factor&\cr
height 2pt&\omit&&\omit&\cr
\noalign{\hrule\vskip3pt\hrule}
height 2pt&\omit&&\omit&\cr
height 12pt depth 5pt&If all the $m_{jk},~j=1,2,3,$ are distinct&&1&\cr
height 12pt depth 5pt&If precisely two of the $m_{jk}$ are distinct&&0&\cr
height 12pt depth 5pt&If all three are equal&&$\;\,\ph_k$&\cr
height 5pt&\omit&&\omit&\cr
\noalign{\hrule}}}
$$\nobreak
\tablecaption{conditions.Z}{The factor for the Yukawa coupling
for each value of $k$.}
\vskip10pt}
\noindent
As an example of the application of the rules of the Table,
we see that
$$
f_{mnp}^3=\ph_1\ph_2\ph_3~e_1e_2e_3 \quad {\rm and} \quad
f_{111}f_{122}f_{133}=\ph_1~e_1e_2e_3~.$$

In order to examine the corrections to the large radius limit and to
make contact with previous work we need to change our parametrization
from the three $\ph_i$ to three flat coordinates $\t_i$,
whose imaginary parts may be thought of as the three ``radii".
The relation between the $\t_i$ and the $\ph_i$ is given by
$$\eqalign{
{}&\ph_i = \ph(\t_i) \cropen{10pt}
{\ph^3\over 4^3}&{(\ph^3 + 8)^3\over (\ph^3 - 1)^3}
 = J(\t) \cr} \eqlabel{Jofphi} $$
with $J(\t)$ the absolute modular invariant of automorphic function theory \
\Ref{\Erdelyi}{A.~Erd\'elyi, F.~Oberhettinger, W.~Magnus and
F.~G.~Tricomi, {\sl Higher Transcendental Functions, Vol. 3},
McGraw--Hill 1953.}.
The relation \eqref{Jofphi} is the same relation as that for a
one-dimensional torus \T\ presented as a cubic in \cp2
$$ x_1^3 + x_2^3 + x_3^3 - 3 \ph\, x_1 x_2 x_3 = 0 $$
and the usual $\t$-parameter.  This will be explained in
Section~\chapref{Zperiods}.

If we denote by $\hat{e}_i$ the monomials corresponding to the $\t_i$,
$$
\hat{e}_i\define -\pd{p}{\t_i}=\ph_i'e_i~,$$
then in the orbifold limit
$$
f_{mnr}^3={\ph_1\ph_2\ph_3 \over
\ph_1'\ph_2'\ph_3'}\,\hat{e}_1\hat{e}_2\hat{e}_3 ~.$$
In the limit that $\t \rightarrow i \,\infty$ the asymptotic form
of $\ph$ is $\ph \sim {e^{-2 \p i \t /3} \over 3}$
so we find that this coupling has a finite nonzero limit as
all $\t_j\to i\,\infty$.  All the other couplings tend to zero in this limit,
and hence we recover the topological couplings \eqref{top}.
The new couplings, however, also contain the instanton corrections
to the topological couplings.
In order to make a detailed comparison with known results we must first
settle some issues involving the choice of basis and the normalization
of the couplings.  We do this in
Section~\chapref{Zperiods},
and we return to the couplings in Section~\chapref{ZYukawanorm}.
The final results are given in Table~\tabref{conditions.cf}
and do indeed agree with the results of Hamidi and Vafa \
\cite\HV.
\bigskip
\subsection
{The Yukawa Couplings for One \hbox{\elevenboldmath \char '163} Nonzero}
\noindent
To calculate the couplings with one $s$ nonzero, we use
the polynomial
$$ p = \sum x_{ij}^3 - 3\, \phi_k \,e_k - 3\, s \,f~,$$
with only one $f$ (so we may take, for example, $f=x_{11} x_{21} x_{31}$).
We find that $e_i^2$ no longer vanishes;
rather we find the relation
$$e_i^2 = {s \phi_i^2 \over 1-\phi_i^3} \,e_i f~.\eqlabel{esquare}$$
Note that this relation is exact, \ie valid to all orders in $s$.

The essential results for the couplings are:
$$\eqalign{
e_i^3 =&
{s^3 \ph_i^3 \ph_1 \ph_2 \ph_3
\Big((1-\ph_i^3)^2(1-\ph_j^3)(1-\ph_k^3) -s^6\Big) \over
  \Big(1-\ph_i^3\Big)^2\Big(1-\ph_i^3 -s^3\Big)
  \Big( (1-\ph_i^3)(1-\ph_j^3) -s^3\Big)
  \Big( (1-\ph_i^3)(1-\ph_k^3) -s^3\Big)}
\,e_1 e_2 e_3\cropen{8pt}
e_i^2 e_j =&
   {s^3 \ph_i^2 \ph_k \over
   \Big(1-\ph_i^3\Big) \Big( (1-\ph_i^3)(1-\ph_j^3)-s^3\Big)}
   \,e_1 e_2 e_3\cropen{8pt}
f^3 =&  {\ph_1 \ph_2 \ph_3 \over 1-s^3 }
\Bigg\{ 1 + {s^3\over 1-\ph_1^3-s^3} + {s^3\over 1-\ph_2^3-s^3}  +
{s^3\over 1-\ph_3^3-s^3} \cr
& \hphantom{{\ph_1 \ph_2 \ph_3 \over 1-s^3 }\Bigg\{ 1 }
+ {s^3\over (1-\ph_1^3)(1-\ph_2^3)-s^3}
  \left(1 + {s^3\over 1-\ph_1^3-s^3} + {s^3\over 1-\ph_2^3-s^3}\right)\cr
& \hphantom{{\ph_1 \ph_2 \ph_3 \over 1-s^3 }\Bigg\{ 1 }
+ {s^3\over (1-\ph_1^3)(1-\ph_3^3)-s^3}
  \left(1 + {s^3\over 1-\ph_1^3-s^3} + {s^3\over 1-\ph_3^3-s^3}\right)\cr
& \hphantom{{\ph_1 \ph_2 \ph_3 \over 1-s^3 }\Bigg\{ 1 }
+ {s^3\over (1-\ph_2^3)(1-\ph_3^3)-s^3}
  \left(1 + {s^3\over 1-\ph_2^3-s^3} + {s^3\over 1-\ph_3^3-s^3}\right)
\Bigg\} \, e_1 e_2 e_3 ~,\cr} $$
from which the remaining couplings,
$e_i e_j f$, $e_i^2 f$, and $e_i f^2$ may be obtained via \eqref{esquare}.

The procedure for obtaining these results is to use the ideal
generated by ${\partial p \over \partial x_{ij}}=0$
to find relations among the products of monomials.
Since these equations are
$$ x_{ij}^2 = \ph_i\, x_{i,j+1} x_{i,j+2}
 + \d_{1j} \,s \,x_{i+1,j} x_{i+2,j} $$
we see that one cannot simplify the calculation
by considering the two indices separately as was the case
in the orbifold limit.
Regardless, a dogged application of these relations
leads first to the discovery of
relationships among the Yukawa couplings such as:
$$\eqalign{
(1-s^3)\,f^3 &= \ph_1 \ph_2 \ph_3 \,e_1 e_2 e_3
+ s \ph_1 \ph_2 \,e_1 e_2 f + s \ph_1 \ph_3\,e_1 e_3 f
 +s \ph_2 \ph_3 \,e_2 e_3 f\cr
&\hskip40pt
+ s^2 \ph_1\, e_1 f^2 + s^2 \ph_2 \,e_2 f^2 + s^2 \ph_3 \,e_3 f^2 \cropen{5pt}
(1-\ph_i^3-s^3)\,e_i f^2 &= s^2 \ph_j\, e_i e_j f + s^2 \ph_k \,e_i e_k f +
s \ph_j \ph_k\, e_1 e_2 e_3 \cr}$$
where $i$, $j$, and $k$ are distinct
and finally to the ability to express all couplings in terms
of a single one, which we have taken to be $e_1 e_2 e_3$.
%
\section{Zmodular}{The Modular Group}
The modular group consists of those transformations of the
periods under which the theory is invariant.
The modular transformations for orbifolds are a well-studied subject \
\REF{\modular}{V.~P.~Nair, A.~Shapere, A.~Strominger and F.~Wilczek,
\npb{287} (1987) 402.\hfill\break
A.~Shapere and F.~Wilczek, \npb{320} (1989) 669.\hfill\break
A.~Giveon, E.~Rabinovici and G.~Veneziano, \npb{322}
(1989) 167.\hfill\break
M.~Dine, P.~Huet and N.~Seiberg, \npb{322} (1989) 301.\hfill\break
W.~Lerche, D.~L\"ust and N.~Warner, \plb{231} (1989) 417.}
\cite{{\LMN,\modular}}.
For the \Z\ orbifold the modular group is known to be
${SU(3,3) \over
   SU(3, \hbox{{\scriptss Z\kern-.4em Z}} )}$ \
\REFS{\FLT}{S.~Ferrara, D.~L\"ust and S.~Theisen, \plb{233} (1989) 147.}
\REFSCON{\Fre}{S.~Ferrara, P.~Fre and P.~Soriani,
Class. Quant. Grav. {\bf 9} (1992) 1649.}\refsend.
For the case of the $\Z/\IZ_3$ orbifold the modular group has been studied by
Shevitz \
\Ref{\Shevitz}{D.~Shevitz, \npb{338} (1990) 283.}\
who has shown the the modular group is
\hbox{$SL(2,\IZ)\otimes SL(2,\IZ)\otimes SL(2,\IZ)$}
each $SL(2,\IZ)$ acting in a familiar way on parameters
$\t_i$, $i=1,2,3$.

It is an interesting question whether this group is preserved
when the singularities of the orbifold are resolved, \ie
the $s_{mnp}$ become nonzero.
We will not settle this issue here.
On first undertaking this investigation our expectation was
that this would not be the case;  however, it appears that the
parameters $s_{mnp}$ are able to accomodate the effect of a
modular transformation on the $\t$'s by becoming
automorphic functions.
Our primary concern here is to use the modular transformations
to select a basis for the periods.
When $s$ becomes nonzero the number of periods increases from
eight to ten and we need to find a new period $z$ and its dual
$\pd{\ca{G}}{z}$ in order to accomodate the two new periods.
This we do by demanding that the modular transformations act
on the period vector via integral matrices.
We again begin our discussion by setting $s_{mnp}=0$.

When the $s_{mnp}$ vanish the polynomial $p$ separates into the sum
$p_1 + p_2 + p_3$ with each $p_i$ corresponding to one torus.
The analysis is therefore largely similar to that for each torus
separately.  There are however some subtleties that prevent the
analogy from being complete.

It is clear that the replacement
$$\ca{A}_i:\ph_i \rightarrow \o \ph_i $$
is a modular transformation since it can be undone by a coordinate
transformation that multiplies one of the coordinates by $\o^{-1}$.
Another that is less obvious but is nevertheless well-known is
$$\ca{B}_i: \ph_i \rightarrow {\ph_i+2\over{\ph_i-1}}~.$$
This also can be undone by a coordinate transformation
$$x_{im}=\l_i \sum_n \o^{mn}\hat{x}_{in}~,$$
with
$$\l_i \define -3^{-{1\over 3}}(\ph_i-1)^{-{1\over 3}}~.\eqlabel{lambda}$$
If we were dealing with just one torus we would not need to correct
an overall scale by means of the factor $\l_i$ but since our
polynomial is in fact $p_1 + p_2 + p_3$ we do have to
allow for this factor.  The operations
$\ca{A}_i$ and $\ca{B}_i$ have the properties that
$\ca{A}_i^3=1$, $\ca{B}_i^2=1$, and between them they
generate the tetrahedral group.

It is instructive to introduce a variable $\g$ defined by
$$\eqalign{
\g(\ph)&=i\left({Z_1(\ph) - \o^2 Z_2(\ph) \over
                 Z_1(\ph) + \o^2 Z_2(\ph)}\right)\cropen{5pt}
  &=2\sqrt{3}\left(\t(\ph) - \half\right)~.\cr}$$
For the moment we simplify the notation by
dropping the subscript which tells us which torus we are working on.
Let \ca{C} be the operation that transports $\ph$ around the branch point at
$\ph=1$. Then \ca{A} and \ca{C} together act on the upper half $\g$--plane
and generate the triangle group
corresponding to the angles $(0,0,\p/3)$.
In fact we have a
representation by matrices given by
$$
\hbox{\ss{A}}=
\pmatrix{\- \cos{\p\over 3}& \sin{\p\over 3}\cr\noalign{\vskip3pt}
     -\sin{\p\over 3}& \cos{\p\over 3}\cr}~~,~~~~
\hbox{\ss{C}\ss{A}}=
\pmatrix{1    & 2\tan {2\p\over 3}\cr\noalign{\vskip3pt}
                        0           &                   1\cr}~.$$

Adding the operation \ca{B} to \ca{A} and \ca{C} gives us the group
${\rm PSL}(2,\IZ)$ with the operations having the
actions
$$
\ca{A}\t={\t-1\over 3\t-2}~~~,~~~~~\ca{B}\t={3\t-2\over5\t-3}~~~,~~~~~
\ca{C}\t=-{\t\over 3\t-1}~.$$
Verifying the action of \ca{B} requires use of the remarkable identities
$$\eqalign{
Z_1\left({\ph+2\over \ph-1}\right)&=
           -{1\over 3}(\ph -1)(2Z_1(\ph)+\hphantom{2}Z_2(\ph))\cropen{5pt}
Z_2\left({\ph+2\over \ph-1}\right)&=
         \- {1\over 3}(\ph -1)(\hphantom{2}Z_1(\ph)+2Z_2(\ph))~.\cr}
$$
The standard ${\rm SL}(2,\IZ)$ generators
$$
\ca{S}:\t\to -{1\over\t}~~,~~~~\ca{T}:\t\to \t+1$$ are given in terms of
\ca{A}, \ca{B} and \ca{C} by the relations
$$
\ca{S}= \ca{B}\ca{A}^2\ca{C}\ca{A}\ca{C}~~,~~~~\ca{T}=
\ca{C}^{-1}\ca{A}^{-1}~.$$

It may be shown that $\ph$ is related to $\t$ by
$$
J(\t) = {\ph^3\over 4^3}{(\ph^3 + 8)^3\over (\ph^3 - 1)^3}~.\eqlabel{ZJofphi}
$$
This relation has 12 branches since if $\ph$ is one branch
then $$
\ca{A} \ph = \o\ph$$
and
$$
\ca{B} \ph = {\ph + 2 \over \ph - 1}$$
are others and successive applications of $\ca{A}$ and $\ca{B}$ give the other
branches.
For definiteness we choose a particular branch; we require $\ph\to\infty$
as $\t\to i\infty$. This still leaves us with a phase ambiguity $\ph\to\o\ph$
but we know that
$$
J\asymp {1\over 12^3 q}~~;~~q\define e^{2\p i\t}$$
as $\t\to\infty$. So we fix the phase by requiring that
$$
\ph\asymp {1\over 3 q^{1/3}}~~\hbox{as}~~\t\to i\infty~.$$

Following \ \cite\LMN\ and \ \cite\CMLN\
we introduce the characters of
the level 1 $SU(3)$ Kac-Moody algebra $\ch_i(\t)$
$$
\ch_i(\t) = {1\over \eta^2(\t)}\sum_{{\bf v}\in\G_i} q^{\half |{\bf v}|^2}$$
where $\eta$ is the Dedekind function
and the $\G_i$ denote the conjugacy classes of the $SU(3)$ weight lattice
($\G_0$ is the root lattice and $\G_{1,2}$ are $\G_0$ shifted by
the fundamental dominant weights).
Note that in fact $\ch_1=\ch_2$. On writing out the sums we find
$$\eqalign{
\eta^2 \ch_0 &=\hphantom{q^{1/3}} \sum_{m,n} q^{(m^2 + mn + n^2)}\cr
\eta^2 \ch_1 &= q^{1/3} \sum_{m,n} q^{(m^2 + mn + n^2 + m + n)}~.\cr}
\eqlabel{Zchiexpand}$$
In~\cite\LMN\  it was shown that under the modular transformation
$\t\rightarrow -1/\t$ the $\ch_i$ transform according to the rule
$$
\pmatrix{\ch_0\cr \ch_1\cr \ch_2\cr}\!\left(-1/\t\right) =
{1\over\sqrt{3}}\pmatrix{1&1&1\cr 1&\o&\o^2\cr 1&\o^2&\o\cr}
\pmatrix{\ch_0\cr \ch_1\cr \ch_2\cr}\! (\t)~. \eqlabel{Z1/tau}$$
It can be shown that
$$
\ph(\t) = {\ch_0(\t)\over\ch_1(\t)} \eqlabel{Zphioftau}$$
in virtue of \eqref{ZJofphi}.
We observe that we are
on the correct branch since $\eta^2\ch_0\to 1$ as
$\t\to i\infty$ while $\eta^2\ch_1\asymp 3q^{1/3}$,
the factor of 3 coming from the three terms
for which
$(m,n)=(0,0),\ \ (-1,0),\ \ \hbox{and}\ \ (0,-1)$.

{}From Equation \eqref{Zphioftau} we have
$$
\ph(\t+1)= \o^2 \ph(\t)$$
and
$$
\ph\left(-1/\t\right) = {\ph(\t) + 2\over \ph(\t) - 1}~.$$
The inverse relation to Equation \eqref{Zphioftau} is
$$\eqalign{
\t(\ph) &= -{i\o\over\sqrt{3}} \left({Z_1(\ph) - Z_2(\ph) \over
Z_1(\ph) + \o^2 Z_2(\ph)}\right)\cropen{10pt}
&={i \over 2 \pi}\left\{ \log(\ph^3) +
{\displaystyle \sum_{n=0}^\infty
  {\txt \G(\third 1{+}n)\G(\third 2{+}n)\over \txt (n!)^2\,\ph^{3n}}
  \Big(2 \Psi(1{+}n) {-} \Psi(\third 1 {+} n) {-} \Psi(\third 2 {+} n) \Big)
\over
\displaystyle\sum_{n=0}^\infty
  {\txt\G(\third 1{+}n)\G(\third 2{+}n)\over \txt (n!)^2\,\ph^{3n}}}
\right\}~.\cr}
\eqlabel{Ztauofphi}$$
which holds for $0<\real\,\t<1$ and the second equality being true for
sufficiently large $\ph$.
\subsection{\hbox{\elevenboldmath \char '163
  \hskip 2pt\elevenboldrm \char '075
  \hskip -9pt \elevenboldrm \char '057
  \hskip 5pt \elevenboldrm 0}
and the Action of the Modular Group on
\hbox{\tenboldmath \char '026} and \hbox{\tenboldmath \char '160}}
\noindent
We proceed to study the action of the modular transformations
on the differential form $\m$ that was used to construct
the holomorphic (5,2)--form and the polynomial $p$,
and we now allow $s_{mnp}$ to be nonzero.
Consider first the coordinate transformation
$$
x_{11}=\o^2\tilde{x}_{11}~\quad,~\quad
x_{im}=\tilde{x}_{im}~~,~~{\rm for}~~(i,m)\ne(1,1)
{}~.\eqlabel{Zwtransf}$$
If we redefine the parameters appropriately
$$
\tilde{\ph}_1=\o^2\ph_1~~,~~\tilde{s}_{1nr}=\o^2s_{1nr}$$
with the other parameters unchanged then the polynomial is invariant in the
sense that
\hbox{$p(\tilde{x}|\tilde{\ph},\tilde{s})=p(x|\ph,s)$}.
This is just a restatement
of the fact that $(\tilde{\ph},\tilde{s})$ and $(\ph,s)$ define the same
manifold and we know from our previous discussion that this change corresponds
to the modular transformation $\t_1\to\t_1+1$. Under this transformation
$\m ={1\over 8!}\e_{A_1 A_2 ...A_9} x^{A_1} dx^{A_2} \ldots dx^{A_9}$ is not
invariant.  In fact $\tilde{\m}=\o\m$.
The modular transformation $\t_1\to-1/\t_1$ corresponds to the more complicated
coordinate transformation
$$x_{1m}=\l_1\sum_n \o^{mn}\hat{x}_{1n}~,\eqlabel{Zsumtransf}$$
with $\l_1$ as in \eqref{lambda}.
These transformations of the parameters are summarized in
Table~\tabref{transf.Z}.

\vbox{
\vskip 10pt
$$
\vbox{\offinterlineskip\halign{
&\vrule#&\strut\quad$#$\quad\hfil
&\vrule#&\quad $#$\hfil\quad
&\vrule#&\quad $#$\hfil\quad
&\vrule#&\quad $#$\hfil\quad
&\vrule#&\quad $#$\hfil\quad
&\vrule#\cr
\noalign{\hrule}
height 14pt depth 10pt&\hfil{\rm Action }\hfil&&\hfil\t_1&&\hfil\ph_1&&\hfil
s_{1np}&&\hfil\m&\cr
\noalign{\hrule\vskip3pt\hrule}
height 16pt depth 10pt&\hfil {\ca T}_1\hfil &&\t_1+1
&&\o^2\ph_1&&\hfil \o^2s_{1np}\hfil &&\hfil \o\m \hfil &\cr
\noalign{\hrule}
height 20pt depth 15pt
&\hfil {\ca S}_1\hfil
&&-{1\over\t_1}
&&{\ph_1 + 2\over \ph_1 - 1}
&&\l_1\sum_k \o^{mk}s_{knp}
&&-3^{-\half}i(\ph_1 - 1)\m
&\cr
\noalign{\hrule}}}
$$
\nobreak\tablecaption{transf.Z}{The transformation of the parameters
under $\ca{T}_1$ and $\ca{S}_1$.}
\vskip 10pt}

We can simplify these transfomation rules by a redefinition of variables.
Consider first the transformation rule for $s_{mnp}$ under
$\t_1\to -1/\t_1$.  This
involves the awkward factor of $\l_1=-3^{-{1\over 3}}(\ph_1-1)^{-{1\over 3}}$.
Notice that under $\ca{B}_1$
$$
\left(\eta^2\ch_1 {d\ph_1\over dJ}\right)^{1\over 3} \rightarrow
-{3^{1\over 6}\over (\ph_1 - 1)^{1\over 3}}
\left(\eta^2\ch_1 {d\ph_1\over dJ}\right)^{1\over 3}~. \eqlabel{Zsfn}$$
So if we set
$$
s_{mnp}=\left(\eta^2\ch_1 {d\ph_1\over dJ}\right)^{1\over 3}v_{mnp}$$
we find that $v$ transforms according to the much simpler rule
$$
v_{mnp}\rightarrow{1\over\sqrt{3}}\sum_k \o^{mk} v_{knp}~.$$
The question arises as to what is meant by the cube roots that appear in
\eqref{Zsfn} and whether they can be consistently defined. The answer is
that the right hand side of \eqref{Zsfn} is defined only up to
multiplication by
cube roots of unity. However this is of no consequence since {\sl all\/} the
$v_{mnp}$ are multiplied by a common phase which can be absorbed by a
simultaneous shift $\t_i\to\t_i + k$. This leaves $\m$ invariant. Note that
$v_{mnp}$ does not return to itself under $\ca{B}_1^2$
but rather to $v_{-m,n,p}$. Thus iterating
the operation $\t_1\to -1/\t_1$ induces
$x_{1m}\to x_{1,-m}$ which changes the sign of $\m$.

We can also
absorb the $\ph_1$ dependent factor that appears in the transformation rule for
$\m$ by noting that under $\ca{B}_1$
$$\left( {d\ph_1\over dJ} \right)^\half \rightarrow
{3^\half i\over (\ph_1 -1)} \left( {d\ph_1\over dJ} \right)^\half~.
\eqlabel{Zmufn}$$
Thus
$$
m\define J^{-{1\over 3}}\left( {d\ph_1\over dJ} \right)^\half \m$$
is invariant under $\t_1\to-1/\t_1$ up to a sign ambiguity introduced by the
square roots in \eqref{Zmufn} ($J^{-{1\over 3}}$, on the other hand, can be
unambiguously defined in virtue of \eqref{ZJofphi}). It is also invariant under
$\t_1\to\t_1+1$ since the factor $J^{-{1\over 3}}$ acquires a factor $\o$.
Note also that under $\t_1\to\t_1+1$ the factor
$\left(\ch_1 {d\ph_1\over dJ}\right)^{1\over 3}$
associated with $v$ is invariant.

The burden of these
observations is that ${m\over p^3}$ can be rendered invariant under modular
transformations up to a sign ambiguity. However the sign ambiguity is of no
consequence for the prepotential or the Yukawa couplings since these have
charge two.

Note that $\ca{B}_i$ is a symmetry only in the limit where all $s_{mnp}=0$,
or if we take a certain subset of the $s_{mnp}$'s.
For example, if $s_{mnp}$ is nonzero then $s_{m+1\,n\,p}$ and $s_{m+2\,n\,p}$
must also be non zero in order for the defining polynomial to be
invariant under $\ca{B}_1$ since
$\ca{B}_1(s_{mnp}) = \l \sum_k \o^{mk} s_{knp}.$
The action of the $\ca{B}_i$ on the $s_{mnp}$ is to
transform an $s$ into a linear combination of $s$'s so this action
will not in general preserve a hypersurface of the parameter space
corresponding to setting some but not all of the $s$'s to zero.

A complete discussion of the modular group would entail letting all
$s_{mnp}$ be nonzero, and investigating the transformation
properties of the flat coordinate $\rho$ associated with each $s$.
The interested reader should consult Reference~\cite\FLT .
%
\section{Zperiods}{The Periods}
The complex structure of a
generalized \cym\ \M\ of, say, seven dimensions
can be described by giving the periods of the holomorphic
form, $\O_{5,2}$, over a canonical homology basis.
By virtue of the Hodge decomposition displayed in
Section~\chapref{Zmanifolds}, we see that the dimension of $H_7(\M,\IZ)$ is
$2(b_{4,3}+1)$.
We proceed in a familiar way by choosing a symplectic basis
$(A^a,B_b)$, $a,b=1,\ldots,b_{4,3}+1$
for $H_7(\M,\IZ)$ such that
$$
  A^a\cap B_b=\d^a{}_b~~,~~A^a\cap A^b=0~~,~~B_a\cap B_b=0~,
  \eqlabel{Zintersect.nos.}$$
and we denote by $(\a_a,\b^b)$ the cohomology basis dual to this so that
$$
  \int_{A^a}\a_b=\d^a{}_b~~,~~\int_{B_a}\b^b=\d_a{}^b~,$$
with other integrals vanishing.
Being a seven--form, $\O_{5,2}$ may be expanded in terms of the basis
$$
  \O_{5,2} = z^a\a_a-\ca{G}_a\b^a~.$$
The coefficients $(z^a,\ca{G}_b)$ are the periods of $\O_{5,2}$,
and are given by the integrals of $\O_{5,2}$
over the homology basis
$$
  z^a=\int_{A^a}\O_{5,2}~\quad,~\quad\ca{G}_b=\int_{B_b}\O_{5,2}~.
\eqlabel{periodint}$$

A comment is in order about the holomorphicity of $\O_{5,2}$.
Let $t$ be one of the $b_{4,3}$ parameters on which the complex
structure depends, and denote by $x^\m$ the coordinates of $\M$,
then it is a fundamental observation in the theory of variation of
complex structure \
\Ref{\Kodaira}{K.~Kodaira, {\sl Complex Manifolds and Deformation
of Complex Structures}, Springer-Verlag, 1986.}\
that
$$ { \partial (\deriv x^\m) \over \partial t}$$
is a linear combination of one-forms of type (1,0) and (0,1) and that
$$  { \partial (\deriv x^\m) \over \partial \bar t} = 0~. \eqlabel{mutbar}$$
It follows from \eqref{mutbar} that
$${ \partial \O \over \partial \bar t} \in H^{5,2} ~.$$
There could, in principle, be a part in $H^{6,1}$ but $H^{6,1}$
is trivial.  In other words,
$${ \partial \O \over \partial \bar t} = h(t, \bar t) \O $$
for some function of the parameters $h$.
Thus by redefining $\O$ by multiplication by a function of the parameters
$$  \O \rightarrow \O \exp
  \left( -\int^{\bar t} h(t,\bar s)\deriv \bar s \right) $$
we can ensure that
$${ \partial \O \over \partial \bar t} = 0 ~. $$
Thus we can assume that $\O$ varies holomorphically with the parameters.
It is now apparent that the formalism of special geometry can
be applied.  We know that the complex structure depends on $b_{4,3}$
parameters so we may take the $\ca{G}_a$ to be functions of the $z^a$.
The $z^a$ are homogeneous coordinates on the space of complex structures
and we have that
$$ \O(\l z) = \l \O(z) ~\quad {\rm and} \quad ~
\ca{G}_b (\l z) = \l \ca{G}_b (z) $$
in the familiar way.  We also observe that
$$ { \partial \O \over \partial z^a} \in
     H^{5,2} \oplus H^{4,3} $$
so we have the relation
$$ \int \O \wedge { \partial \O \over \partial z^a} = 0 $$
which has the consequence that the $\ca{G}_a$ are the derivatives of
a prepotential $\ca{G}$ of homogeneity degree two:
$$ \ca{G}_a = {\partial \ca{G} \over \partial z^a} ~. $$

The multiplication of $\O$ by a holomorphic function of the moduli
$$ \O \rightarrow f\,\O \eqlabel{omegagauge}$$
has no effect on the metric on moduli space or on the invariant
Yukawa couplings.  However, it has a direct effect on the periods
and their modular transformations.
We shall choose a particular gauge in Section~\chapref{Zperiods}.5 in order
to find an integral basis for the periods.

The form $\O$ and its derivatives with respect to the complex structure
parameters are all seven-forms.
There are at most $2(b_{4,3}+1)$ linearly independent such quantities.
It follows that $\O$ satisfies linear differential equations:
the Picard-Fuchs equations.
For the mirror of the \Z\ manifold these provide a
straightforward way of computing the periods.
Derivations of the Picard-Fuchs equations
intended for physicists are given in \
\REFS{\CF}{A.~C.~Cadavid and S.~Ferrara, \plb{267} (1991) 193.}
\REFSCON{\BV}{B.~Blok and A.~Varchenko, Int. J. Mod. Phys. A{\bf 7},
(1992) 1467.}
\REFSCON{\LSW}{W.~Lerche, D.~J.~Smit and N.~P.~Warner, \npb{372} (1992) 87.}
\refsend,
the method having derived its origin in the work of Griffiths \
\Ref{\Griffiths}{P.~Griffiths, Ann. Math. {\bf 90} (1969) 460.}.

We shall employ this method to find the periods in the orbifold limit
when all $s_{mnp}=0$.  We relate this to a set of periods calculated
by choosing a homology basis and directly evaluating the periods
via \eqref{periodint}.  We also exploit the relationship of
\Ztilde\ in the orbifold limit to a product of tori:
the homology of \Ztilde\ factorizes, and so we use our understanding
of the homology of the torus to define a basis satisfying
\eqref{Zintersect.nos.}.  Thus we find an integral basis for the
periods when all $s_{mnp}=0$.

When we allow one $s$ to be nonzero the homology no longer
factorizes.  To determine a suitable basis for the periods we
insist that the modular group act on the period vector by integral
matrices.  We find first a basis of periods that satisfy the Picard-Fuchs
equations and which reduce to the periods of the integral basis in the orbifold
limit. The next step is to investigate the effect of modular transformations on
the periods. We do this in Section~\chapref{Zperiods}.5.
\subsection{The Periods on the Mirror when All
\hbox{\elevenboldmath \char '163 \lower.5ex\hbox{\eightboldmath \char '155
 \eightboldmath \char '156 \eightboldmath \char '160}
  \hskip 2pt\elevenboldrm \char '075\hskip 2pt \elevenboldrm 0}}
\noindent
When all $s_{mnp}$ vanish we expect 8 independent periods.
In this limit the defining polynomial is
$$
p=\sum _{i=1}^9 y_{i}^3 - 3\sum _{k=1}^3 \ph_k e_k ~,\eqlabel{Zp}$$
which is suggestively like three copies of
the equation used to define a torus as a \cp2[3] with a polynomial
$$
p_1=x^3 + y^3 + z^3 - 3\, \ph\, x y z  ~.\eqlabel{torusp}$$
Both \cite\CF\ and \cite\LSW\ derive the differential equation
satisfied by the periods for a torus;
they find that each period $q$ is a solution of the equation
$$
\left[ \ph^3\left(1-\ph^3\right){d^2\over d\left(\ph^3\right)^2} +
        \left({2\over 3} - {5\over 3}\ph^3\right) {d\over d\ph^3}
   -{1\over 9} \right] \, q = 0~.\eqlabel{Zhge}$$
The Picard-Fuchs equations for \Ztilde\ turns out to be
just three copies of this, with $\ph$ replaced by $\ph_i$.
We can show directly that the periods on the manifold must solve
Equation~\eqref{Zhge} for each $\ph_i$ by following the procedure
outlined in \cite\LSW, using $\ph_1$ as an example.
A period $q$ is an integral of $\O$ \eqref{res} over a seven-cycle,
so we write
$$q = \int {\m \over p^3}~.$$
Then two derivatives with repect to $\ph_1$ yields
$$\eqalign{
{\partial^2 q \over \partial \ph^2_1} &= 4 \cdot 3^3 \int \m
{(y_1 y_2 y_3 )^2  \over p^5}\cropen{5pt}
&=4 \cdot 3^3 \int \m  {(y_1 y_2)^2  \over p^5}
\left( {1 \over 3}{\partial p \over \partial y_3} +
    \ph_1 \,y_1 y_2 \right) \cropen{5pt}
&=4 \cdot 3^3 \int \m \left( {\ph_1 (y_1 y_2)^3  \over p^5} -
{1 \over 12}{\partial \over \partial y_3 }
\left( {y_1^2 y_2^2 \over p^4}\right) \right)~.\cr}$$
In the second line we have used
$y_3^2 = \third{1} {\partial p \over \partial y_3} + \ph_1 y_1 y_2$.
The second term of the third line is zero.
We then make similar substitutions in the first
term, first with
${\partial p \over \partial y_2}$, then with
${\partial p \over \partial y_1}$,
and set to zero all integrals of total derivatives.  The final result is that
$$q'' = \ph_1 q + 3 \ph_1^2 q' + \ph_1^3 q''~,$$
where $q' = {\partial  q \over \partial \ph_1}$.  If we change variables to
$\ph_1^3$ we find Equation~\eqref{Zhge}.
Thus the periods on \Ztilde\ must solve this equation for each $i$.
This is a surprise on first aquaintance since our period $q$ contains
an integral over ${1\over p^3}$ rather than the integrals over
${1\over p_1}$ that arise for the torus.
The reason that the periods nevertheless satisfy the same equations
is that the homology is carried by the three tori $\T_i$.
We shall explain this presently.

The Equation~\eqref{Zhge} is of course a hypergeometric equation
for which the solutions can be represented by a Riemann P--symbol
$$
\Psymbol{0&{1\over3}&0\cr
\noalign{\vskip3pt}{1\over3}&{1\over3}&0\cr}{\phi^3}~.$$
There are two independent solutions; we define
$$\eqalign{
Z_1(\ph)&=\hphantom{\ph}{\G^2(\third{1})\over\G(\third{2})}
       \F(\third{1},\third{1};\third{2};\ph^3)\cr
Z_2(\ph)&=          \ph {\G^2(\third{2})\over\G(\third{4})}
       \F(\third{2},\third{2};\third{4};\ph^3)\cr}$$
and conclude that there is a basis of 8 independent periods
given by the product functions $Z_i (\ph_1) Z_j(\ph_2) Z_k(\ph_3)$,
where $i$, $j$, and $k$ are independently chosen to be 1 or 2.

Another method of calculating the periods is to proceed by choosing
a basis for $H_7$ and directly integrating $\O$ over these
seven-cycles \eqref{periodint}.  We leave the details for the Appendix;  the
resulting periods factorize, as they must, into products of solutions of
\eqref{Zhge}.

Our solutions are products of solutions
of the Picard-Fuchs equation for a torus.
Hence it is intuitive that we may
calculate a basis by using the homology cycles on the torus.
In this basis we can easily find the intersection of the
cycles, and so it is this basis which we will use to define
the integral periods.  To explain this we turn to a discussion
of the periods on the torus and their relevance
before calculating the periods on \Ztilde\ away from the orbifold point.
\subsection{Three Tori from One Seven--fold}
\noindent
Let us examine why the periods on the mirror of the \Z\ break into products of
three periods on tori in the orbifold limit.
We note that from Equation~\eqref{Zp},
we can break $p$ into a sum of three polynomials $p = p_1 + p_2 + p_3$, where
$$ p_i=\sum _{k=3i-2}^{3i} y_{k}^3 - 3 \ph_i e_i~.$$
For a seven-cycle $H$,
$$ \eqalign{
\int_H \O_{5,2} &= {1\over{2 \p i}}\int_H \int_{C_p}
                       { y_9 dy_1 \ldots dy_8 \over p^3} \cropen{5pt}
      &={1\over(2 \p i)^2}\int_H \int_{C_p} \int_{C_{y_9}}
                        { dy_1 \ldots dy_9 \over p^3}\cropen{5pt}
  &={1\over(2 \p i)^2}\int_{H \times C_p \times C_{y_9}}
                        {dy_1 \ldots dy_9 \over 6p_1 p_2 p_3}
        \sum_{n=0}^\infty ({\D\over6p_1 p_2 p_3})^n~,\cr}$$
where $\D = (p_1 + p_2 + p_3)^3 - 6p_1 p_2 p_3$
and the second equality follows due to the fact that the integral
is actually independent of the value of $y_9$ in virtue
of the homogeneity of the integrand.
We may therefore introduce a factor of ${1\over 2 \p i y_9}$
and integrate around a loop $C_{y_9}$ without changing the
value of the integral.
Suppose our cycle $\tilde H = H \times C_p \times C_{y_9}$ contains the circles
$C_{p_1} \times C_{p_2} \times C_{p_3}$;  then it is easy to see that we get a
contribution from only the zero'th order term in the sum.

The period is proportional to
$$
\int_{\tilde H} {dy_1 \ldots dy_9 \over p_1 ~p_2~ p_3}$$
and we see that the integrand obviously breaks into the product of three
integrands of the form in Equations \eqref{Zs2torus} and \eqref{Zs1torus}
above.
The integral becomes separable when $\tilde H$ is separable.
For example, suppose we take a seven-cycle
$H=\{ C_{y_1}\times\ldots\times C_{y_7}\}$.  Then if we solve $p=0$ for $y_8$,
choosing the branch $y_8 \rightarrow 0$ as $\phi_3\rightarrow \infty$,
we see that we have three copies of the integral in
Equation \eqref{Zs2torus} (provided we are in the regime where all $\phi_i$ are
large).
\subsection{The Periods on a Torus}
\noindent
Consider the torus as a cubic in \cp3 given by \eqref{torusp}
and define the contour
$A_3 = C_{p_1} \times C_x \times C_y$,
where $C_f$ is a contour that winds around the hypersurface $f=0$.
In the limit $\phi \rightarrow \infty$ and
on the branch of $p_1=0$ for which
$z \rightarrow 0$ as $\phi\rightarrow \infty$,
we calculate the period $R$ to be
$$\eqalign{
R &= \int_{A_3} {dx dy dz \over p_1}\cr
    &= -{1 \over 3 \phi} (2 \pi i )^3 \F(\third 1,\third 2;1;\phi^{-3})\cr
    &= {(2\p)^2 \over 3}(\o-1)(Z_1(\ph)+ \o^2 Z_2(\ph))~.\cr}
\eqlabel{Zs2torus}$$
Alternatively, we could choose to integrate ${dy \over dp_1/dz}$ over
the one--cycle
$$\eqalign{
A=&\{x=1,\ C_y, \cr
    &\kern15pt z \hbox{ given by the branch of }p_1=0
\hbox{ such that }z\rightarrow 0\hbox{ as }\phi \rightarrow \infty\}~.\cr}$$
We then find that
$$R = (2 \pi i)^2 \int_{A} {dy \over \partial p_1/\partial z}~.$$

In order to find a second solution,
we should examine $p_1$ as $\phi\rightarrow 1$.
There is a singularity near $(x,y,z)=(1,1,1)$.
In the neighborhood of this singularity we let
$$\eqalign{
x &= 1~,\cr
y &= 1+u_2~,\cr}\hskip40pt
\eqalign{
z &=1+u_3~,\cr
\phi&=1 + \epsilon^2~.\cr}$$
Substituting this into $p_1$ we arrive at
$$
p_1 = u_2^2 + u_3^2 - u_2 u_3 - \epsilon^2~.$$
Thus for $u_2$ and $u_3$ real, there is a circle that shrinks to zero as
$\epsilon\rightarrow 0$.  Since this is a one-cycle, we can integrate a
one--form over it.  If we define
$$\eqalign{
B = &\big\{x=1,\ y = 1 + u_2,\cr
      &\hskip15pt z \hbox{ follows the } S^1
      \hbox{ that shrinks to zero as } \phi\rightarrow 1\big\}~,\cr}$$
then the period $Q$ is
$$\eqalign{
Q &= (2 \pi i)^2 \int_{B}{dz \over \partial p_1/\partial y} \cr
  &= -{(2 \pi)^3 \over 3 \sqrt 3} \F(\third 1,\third 1,1;1-\phi^3)\cr
  &=  {(2 \pi)^2\over3} (-Z_1(\ph) + Z_2(\ph) )~.\cr}
\eqlabel{Zs1torus}$$
We remark that we are integrating the same one form as before, because
${dz \over \partial p_1/\partial y}  = - {dy \over \partial p_1/\partial z}$.
The importance of the periods $Q$ and $R$ lies in their
relation to the $\t$--parameter. The relation is
$$
\t =  {Q\over R}~.$$

Of course, we could have defined
a three cycle to be integrated over a three form as well:
$$ B_3 = \left\{C_x \times C_{p_1} \times
  \hbox{the }S^1 \hbox{ near (1,1,1) that shrinks to zero as }
  \phi\rightarrow 1\right\}~,$$
so
$$Q = \int_{B_3}{dx dy dz \over p_1}~.$$

We know that we can define a period on the 7--dimensional
manifold by a product of these periods.
We arrange the 8 independent periods into a column vector,
let $\vp_{{\scriptscriptstyle QQQ}}$ stand for $Q(\ph_1) Q(\ph_2) Q(\ph_3)$,
{\it etc.}, and define
$$
\Pi_8 =\pmatrix{  -\vp_{{\scriptscriptstyle QQQ}}\cr
               \-  \vp_{{\scriptscriptstyle RQQ}}\cr
               \-  \vp_{{\scriptscriptstyle QRQ}}\cr
               \-  \vp_{{\scriptscriptstyle QQR}}\cr
               \-  \vp_{{\scriptscriptstyle RRR}}\cr
               \-  \vp_{{\scriptscriptstyle QRR}}\cr
               \-  \vp_{{\scriptscriptstyle RQR}}\cr
               \-  \vp_{{\scriptscriptstyle RRQ}}\cr}
= {1\over  R(\ph_1)R(\ph_2)R(\ph_3)}
  \pmatrix{  -\tau_1 \tau_2\tau_3\cr
           \- \tau_2 \tau_3\hfil \cr
           \- \tau_1 \tau_3\hfil \cr
           \- \tau_1 \tau_2\hfil \cropen{3pt}
           \- 1\hfil \cr
           \- \tau_1 \hfil\cr
           \- \tau_2 \hfil\cr
           \- \tau_3 \hfil\cr}
           \-
{}~.\eqlabel{Pieight}$$

It remains to discuss the intersection of the cycles.
Consider again the neighborhood of the singularity.
Locally the singularity approximates a $S^0=\IZ_2$ bundle over an $S^1$.
$S^0$ consists of two points so we have a double cone.
What is happening is that a short cycle on the torus
is being shrunk to a point, and so the torus becomes
a sphere with two points identified.
The basis we have chosen is the short cycle that shrinks
to zero ($B$) and the long cycle that passes through the node ($A$).
It is clear that the cycles $B$ and $A$ intersect in a point.
It is this fact which allows us to define elements of the integral
homology in the seven--fold.
The cycles on the 7--manifold that produce the
periods that are products of the periods $Q$ and $R$ on the tori
have intersection numbers that are just given by the product
of the intersections of the cycles $B$ and $A$ on the tori.
Thus the cycles that produce the periods in \eqref{Pieight}
have intersection numbers ~\eqref{Zintersect.nos.}, and we recover in this way
the basis of Shevitz \ \cite{\Shevitz}.
\subsection{The Periods for One
\hbox{\elevenboldmath \char '163 \lower.5ex\hbox{\eightboldmath \char '155
 \eightboldmath \char '156 \eightboldmath \char '160}} Nonzero}
\nobreak
\noindent
It is of interest to examine the periods on the manifold as a function of the
$s_{mnp}$.
Suppose we let only one $s_{mnp} \equiv s$ be non-zero, and so
we take $$ p=\sum _{i=1}^9 y_{i}^3 - 3\sum _{k=1}^3 \ph_k e_k
    - 3  s f_{mnp}~.$$
The homology no longer factors, but
our method of defining a set of cycles and
integrating to find the periods is still available,
and yields the periods as power series in $s$.
We compute the periods in this way in the Appendix and find,
as we must, that there are now 10 independent periods.
These periods are:
\item{$\bullet$~~}
A set of eight periods of the form
$$
\vp_{ijk}= \sum_{r=0}^\infty {(3s)^{3r}\over (3r)!}
Z_i(\ph_1,r)Z_j(\ph_2,r)Z_k(\ph_3,r)$$
where the functions
$$\eqalign{
Z_1(\phi,r) &=
  {\G\left(\third{1}\right)\G\left(r+\third{1}\right)\over
       \G\left(\third{2}\right)}
\F(\third{1},r + \third{1};\third{2};\phi^3) \cr
Z_2(\phi,r) &=
  {\G\left(\third{2}\right)\G\left(r+\third{2}\right)\over
       \G\left(\third{4}\right)}\phi\,
\F(\third{2},r + \third{2};\third{4};\phi^3) \cr}$$
generalise the functions $Z_1(\ph)$ and $Z_2(\ph)$ defined previously
which correspond to $Z_1(\ph,0)$ and $Z_2(\ph,0)$ respectively. These periods
reduce to the eight periods of the orbifold limit as $s\to 0$.
\item{$\bullet$~~}
A period that is $\ca{O}(s)$ as $s\to 0$
$$
\vp= \sum_{r=0}^\infty {(3s)^{3r+1}\over (3r+1)!}
Z_3(\ph_1,r)Z_3(\ph_2,r)Z_3(\ph_3,r)~~~,~~~
Z_3(\ph,r)\define Z_1(\ph, r+\third{1})~.$$
\item{$\bullet$~~}
A period that is $\ca{O}(s^2)$ as $s\to 0$
$$\widehat{\vp}= \sum_{r=0}^\infty {(3s)^{3r+2}\over (3r+2)!}
Z_4(\ph_1,r)Z_4(\ph_2,r)Z_4(\ph_3,r)~~~,~~~
Z_4(\ph,r)\define Z_2(\ph, r+\third{2})~.$$

\noindent The Picard-Fuchs equations are still relatively simple.
One may use the same method as in the orbifold case to derive the differential
equations satisfied by the holomorphic form $\O$.
As a practical matter, though,
it seems simpler to find the periods by direct integration as in
Appendix~A and then find the differential equations that they
satisfy.

It is easy to see that each of the ten periods above satisfies the equation
$$
\left\{ \ph_a^3 \left(1-\ph_a^3\right)
    {\partial^2 \over \partial \left(\ph_a^3\right)^2} +
        \left( {2\over 3}  - {5\over 3} \ph_a^3 \right)
  {\partial \over \partial \ph_a^3}
- \left({\ph_a^3 \over 3} {\partial \over\partial \ph_a^3}
 +{1\over 9}\right) s {\partial \over \partial s}
 -{1\over 9} \right\} \O= 0~\eqlabel{Zdiffq}$$
for each $a=1,2,3$. A further equation follows from the observation that
$$
Z_i(\ph,r+1)=
\left[ \ph^3{d\over d\ph^3}+\left(r+{1\over 3}\right)\right]Z_i(\ph,r)~~~,
{}~i=1,2,$$
from which it follows that each of the periods satisfies the relation
$$
\left\{ L_1L_2L_3 - \left({1\over 3}\pd{}{s}\right)^3 \right\}\O =0
\eqlabel{PFtwo} $$
with $L_a$ denoting the operators
$$
L_a=\ph_a^3\pd{}{\ph_a^3} + {s\over 3}\pd{}{s} + {1\over 3}~.$$

It remains to extend the integer basis that we had for $s=0$ to include the
two new elements.  Our criterion will be to demand that the
modular transformations act on the basis by means of integral matrices. We may
choose eight of the ten basis elements to be the linear combinations of the
$\vp_{ijk}$ that reduce to the periods \eqref{Pieight} in the limit $s\to 0$.
these are periods $\vp_{{\scriptscriptstyle QQQ}}~,\vp_{{\scriptscriptstyle
QQR}}$ {\it etc.\/} where now $\vp_{{\scriptscriptstyle QQQ}}$, for example,
denotes $\sum_{r = 0}^\infty {(3 s)^{3r}\over (3r)!}Q(\ph_1,r) Q(\ph_2,r)
Q(\ph_3,r)$ and $Q(\ph,r)$ and $R(\ph,r)$ are defined as the obvious
generalizations of the corresponding quantities with $r=0$
$$
\pmatrix{Q(\ph,r) \cropen{3pt} R(\ph,r) \cr}= {(2\p)^2\over 3}
\pmatrix{-1& 1\cropen{3pt} -i\sqrt{3}\o^2& -i\sqrt{3}\o\cr}
\pmatrix{Z_1(\ph,r) \cropen{3pt} Z_2(\ph,r) \cr}~. \eqlabel{QR} $$
\subsection{An Integral Basis for the Periods}
We are finally ready to find an integral, symplectic basis for the periods.
First recall that, by construction, the
period vector $\Pi_8$ of \eqref{Pieight} corresponds to a symplectic basis for
the corresponding cycles and it possesses the right intersection
numbers~\eqref{Zintersect.nos.}.
Thus we may use it to calculate the invariant couplings as described in \
\Ref{\CdGP}{P.~Candelas, X.~de la Ossa, P.~Green and L.~Parkes,
\npb{359} (1991) 21.\newpage}\
and Section~\chapref{ZYukawanorm}.
We would like to show explicitly that there is a gauge
in which the periods transform by integers under the modular group.

The modular group is in this context the group generated by the $\ca{A}_i$ and
the $\ca{C}_i$ since the transformations generated by the $\ca{B}_i$ do not
respect the condition that only one $s$ is nonzero. The generators \ca{A}\ and
\ca{C}\ do not change powers of $s$ so it is clear that we may take eight of
the ten basis elements to correspond to the eight elements of $\Pi_8$.
The issue reduces to how to choose linear combinations of the new periods
$\vp$ and $\widehat{\vp}$ in order to obtain the remaining periods.
The action of \ca{A}\ and \ca{C}\ on the functions $Z_1(\ph,r)$ and
$Z_2(\ph,r)$ is independent of $r$ if $r$ is integral :
$$
\ca{A}~:~\pmatrix{Z_1\cr Z_2\cr} \rightarrow
\pmatrix{1&0\cr 0&\o\cr}\pmatrix{Z_1\cr Z_2\cr}~~~~,~~~~
\ca{C}~:~\pmatrix{Z_1\cr Z_2\cr} \rightarrow
\pmatrix{-2\o& i\sqrt{3}\cr -i\sqrt{3}& -2\o^2\cr}\pmatrix{Z_1\cr Z_2\cr}~.$$
We see, in virtue of \eqref{QR} that
$$
\ca{A}~:~\pmatrix{Q\cr R\cr}\rightarrow
\o^2\pmatrix{1&  -1\cr 3&  -2\cr}\pmatrix{Q\cr R\cr}
{}~~~~,~~~~
\ca{C}~:~\pmatrix{Q\cr R\cr}\rightarrow
\pmatrix{\- 1&\- 0\cr  -3&\- 1\cr}\pmatrix{Q\cr R\cr}
{}~.\eqlabel{AandC}$$
The matrices that appear in this basis are the matrices {\ss{A}}\ and {\ss{C}}
of Section~\chapref{Zmodular}, apart from a factor of automorphy of $\o^2$ that
appears in relation to \ca{A}. It is clear from these matrices that, apart from
this factor of automorphy, the period vector $\Pi_8$ transforms integrally
under \ca{A}\ and \ca{C}. Turning now to the new periods we find that
$$
\ca{A}~:~\pmatrix{\vp\cr \widehat{\vp}\cr}\rightarrow
\o^2\pmatrix{\o^2&0\cr 0&\o\cr}\pmatrix{\vp\cr \widehat{\vp}\cr}~~~~,~~~~
\ca{C}~:~\pmatrix{\vp\cr \widehat{\vp}\cr}\rightarrow
\pmatrix{\o^2&0\cr 0&\o\cr}\pmatrix{\vp\cr \widehat{\vp}\cr}~.$$
Note that the same \hbox{SL(2)} matrix appears in each case and that \ca{A}\ is
accompanied by the same factor of automorphy as in \eqref{AandC}.

We are to find a change of basis
$$
\pmatrix{\ca{G}_4\cropen{3pt} z^4\cr}=
M\pmatrix{\vp\cropen{3pt} \widehat{\vp}\cr}~~~~,~~~~
M=\pmatrix{a&b\cr c&d\cr} \eqlabel{G4z4}$$
such that the matrix that represents \ca{C}\ becomes integral. In \eqref{AandC}
\ca{C}\ is represented by a matrix whose cube is the identity. It is well known
that the the only matrices in $\hbox{Sp(2,\IZ)} = \hbox{SL(2,\IZ)}$
with this property are, up to
conjugation, the matrix $\left({\- 0~\- 1\atop -1~-1}\right)$ and its inverse
$\left({-1~-1\atop\- 1~\- 0}\right)$. Thus we require
$$
M\pmatrix{\o^2&0\cr 0&\o\cr}M^{-1}=\pmatrix{\- 0&\- 1\cr -1&-1\cr}
{}~~~\hbox{or}~~~
M\pmatrix{\o^2&0\cr 0&\o\cr}M^{-1}=\pmatrix{- 1&- 1\cr \- 1&\- 0\cr} ~.$$
A consideration of the the periods
calculated in the Appendix by integrating over combinations of
integral cycles leads us to
$$M=c\,\pmatrix{\o^2&\o\cr 1&1\cr}
{}~,~~\hbox{where}~~~c={(2 \pi)^6 i \over 3^{5/2}}~, \eqlabel{Manda}$$
which satisfies the second equality.
In this way we obtain the ten--component period vector
$$\Pi \define \pmatrix{\ca{G}_0 \cr
                       \ca{G}_1\cr
                       \ca{G}_2 \cr
                       \ca{G}_3 \cr
                       \ca{G}_4 \cropen{3pt}
                            z^0 \cr
                            z^1 \cr
                            z^2 \cr
                            z^3 \cr
                            z^4 \cr}
=   f  \pmatrix{  -\vp_{{\scriptscriptstyle QQQ}}\cr
               \-  \vp_{{\scriptscriptstyle RQQ}}\cr
               \-  \vp_{{\scriptscriptstyle QRQ}}\cr
               \-  \vp_{{\scriptscriptstyle QQR}}\cr
               \-  c(\o^2\vp + \o\widehat{\vp})   \cropen{3pt}
               \-  \vp_{{\scriptscriptstyle RRR}}\cr
               \-  \vp_{{\scriptscriptstyle QRR}}\cr
               \-  \vp_{{\scriptscriptstyle RQR}}\cr
               \-  \vp_{{\scriptscriptstyle RRQ}}\cr
                    c(\vp + \widehat{\vp})       \cr}~.\eqlabel{Piten}$$
Modular transformations induce factors of automorphy in the periods as has been
noted above. These factors are the same as those that we encountered in the
transformation of $\m$ in Section~\chapref{Zmodular}. We have therefore
introduced a gauge factor
$$
f=\prod_{i=1}^3 J^{-\third{1}}(\ph_i) \left({d\ph_i \over dJ}
        \right)^{1/2} ~.\eqlabel{Zfchoice}$$
Thus defined $\P$ transforms with integral symplectic transformations and
without any factors of automorphy.
The reader interested in a further discussion of the modular properties
of the periods is referred to the works of \
\REFS{\FLTtwo}{S.~Ferrara, D.~L\"ust and S.~Theisen, \plb{242} (1990) 39.}
\REFSCON{\Villa}{M.~Villasante, {\sl Solutions to the Duality Equations
for Blown-Up Orbifolds}, \hfil\break UCLA/91/TEP/37.}\refsend.
%
\section{ZYukawanorm}{The Normalized Yukawa Couplings}
We are now able to return to the calculation of the Yukawa
couplings and the metric on the moduli space with a view to comparing with
known results.
In virtue of the relations given by special geometry, the Yukawa couplings are
given by
$$\eqalign{
y_{ABC} &= \int \O \wedge \partial_{ABC} \O~, \cr
  &=\Pi^{{\ss T}} \, \Sigma \, \partial_{ABC} \Pi~,  \cr}\hskip40pt
\Sigma =\pmatrix{0&{\bf 1}\cr{\bf -1}&0\cr}~.$$
The $y_{ABC}$ are dependent
upon the gauge choice \eqref{omegagauge}.  To make the comparison with previous
work we consider the invariant Yukawa coupling (see \ \cite\CdGP)
$$ \k_{ABC} = \rootmetric{A}\, \rootmetric{B}\,\rootmetric{C}\
                e^K \, |y_{ABC}| ~.$$
Here $K$ is the \K\  potential and
$g_{A,\bar{B}}$ the metric on the parameter space which are given
in terms of the period vector by
$$
 e^{-K} = -i\,\Pi^{\dagger} \, \Sigma \, \Pi \hskip40pt
 g_{A,\bar{B} } = \partial_A\partial_{\bar{B}} K ~.$$

The moduli space is $b_{4,3}$--dimensional and the $\ph_i$'s together
with the $s_{mnp}$'s form a set of coordinates on the moduli space.
We can also use ratios of the integral periods as coordinates. These correspond
to the `flat coordinates'\
\Ref{\FL}{S.~Ferrara and J.~Louis, \plb{278} (1992) 240.}.
In these coordinates many expressions take a simple form, as we will see. Some
of the quantities of interest may be obtained by setting $s=0$ {\it ab
initio\/} and working with the periods \eqref{Pieight}. This will not yield all
the Yukawa couplings in the orbifold limit; the coupling
$\k_{sss}$ for example receives contributions from the order $s^3$ terms in the
periods that survive the limit $s\rightarrow 0$. In order to calculate the
metric components $g_{s\bar{s}}$ we must also differentiate the \K\ potential
before setting $s=0$.

We choose flat coordinates:
$$
\z^a = {z^a\over z^0}~,~a=1,2,3,~~~~~\r={z^4\over z^0}~.$$
When $s=0$ the $\z^a$ coincide with the $\t_a$ discussed previously and in this
limit we shall use the $\z$'s and the $\t$'s interchangeably. When $s\ne 0$ we
shall continue to think of the $\t$'s as related to the $\ph$'s by
\eqref{Zphioftau}.

The exponential of the \K\ potential is
$$\eqalign{
e^{-K} &= -i  |f|^2  \prod_{i=1}^3
 \Big( Q(\phi_i)\bar{R(\ph_i)} -R(\phi_i)\bar{Q(\ph_i)}\Big)\cr
&=  -  2^3 |z^0|^2 \prod_{i=1}^3 \imag\,\tau_i ~,\cr} $$
with $f$ chosen as in \eqref{Zfchoice}.
We see from \eqref{Pieight} that the prepotential is given by the expected
quantity \
\REF{\CFG}{S.~Cecotti, S.~Ferrara and L.~Giarardello, \plb{213} (1988) 443.}
\cite{{\CFG,\Shevitz}}
$$
\ca{G} ~=~ {z^1 z^2 z^3 \over z^0} ~=~ (z^0)^2 \,\t_1 \t_2 \t_3~.$$
This leads to the metric components
$$
g_{\t_i, \bar\t_j}  = {\d_{ij} \over 4 (\imag \,\t_i )^2 } ~,$$
which is standard metric on the moduli--space of the torus.
The only nonzero Yukawa couplings are $y_{\t_1,\t_2,\t_3}$ and $y_{sss}$.
We see that
$$
y_{\t_1 \t_2 \t_3 } = (z^0)^2 ~,$$
for which the invariant coupling is seen to be
$$\k_{\t_1 \t_2 \t_3 } =
  \sqrt{g^{\t_1,\bar{\t}_1}}  \sqrt{g^{\t_2,\bar{\t}_2}}
  \sqrt{g^{\t_3,\bar{\t}_3}}\ e^K\ | y_{\t_1 \t_2 \t_3 } | = 1~.$$

In the limit that $s\to 0$ the mixed terms of the metric
$g_{s,\bar{\imath}}$ or $g_{i,\bar{s}}$ vanish, while we find that
$$ \eqalign{
g_{s,\bar{s}} &=(2\p)^{12} 3^{-{5\over 2}}i\, \prod_{i=1}^3
     {|Z_3(\ph_i)|^2 \over
  R(\phi_i)\bar{Q(\ph_i)} -Q(\phi_i)\bar{R(\ph_i)} }~. \cr}
$$
Apart from the coupling
$\k_{\tau_1 \tau_2 \tau_3}$ the only other coupling that survives the limit
$s\to 0$ is
$$\eqalign{
y_{sss} &= 27\, f^2\, \prod_{j=1}^3
\Big(R(\phi_j,0) Q(\phi_j,1) -  Q(\phi_j,0) R(\phi_j,1)\Big) \cr
        & = \phi_1 \phi_2 \phi_3 \, y _{\ph_1 \ph_2 \ph_3}~,\cr}$$
which is in agreement with the calculations in
Section~\chapref{ZYukawas} using the cohomology ring,
{\it cf.} Table~\tabref{conditions.Z}.
Changing variables,
we find that to lowest order in $s$,
$$
y_{\rho\rho\rho} =  (z^0)^2\, \prod_{k=1}^3{3 \sqrt2 \over (2 \pi)^3}
                                      \nu \phi_k R(\phi_k) ~=~
(z^0)^2\,\prod_{k=1}^3 \sqrt2 i\,\nu \eta^2(\t_k) \ch_0(\t_k)~,$$
where  $\nu =\sqrt{{3\over2}}{\G^2(\third{2})\over\G(\third{1})}$
and we have used the relation\Footnote{This relation was suggested by the
analysis of this section. We have checked it numerically but we do not have an
analytic proof.}
$$ R(\phi) = {8 \pi^3 i \over 3} \eta^2 \chi_1~,$$
with $\ph(\tau)$ is given by Equation\ \eqref{Zphioftau}.

The normalized coupling is
$$\eqalign{
\k_{sss} &= (g^{s \bar{s}})^{3\over2}\, e^K\, |y_{sss}|\cr
        &=
\prod_{j=1}^3 {3^{1/4} \over2 \pi \G^3(\third{1})}
    \left|\phi_j \right| \left| R(\phi_j)\bar{Q(\ph_j)} -
          Q(\phi_j)\bar{R(\ph_j)}\right|^{{1\over2}}\cr
&= {3^3 \over (2 \pi)^9} \prod_{j=1}^3 \nu r_j |\ph_j R(\ph_j)| \cr
&= \prod_{j=1}^3 \nu r_j |\eta^2(\t_j) \chi_0(\t_j)| ~,\cr }$$
where $r_j = {2 \over 3^{1/4}}\sqrt{\imag \, \t_j}$.
Now we can write the coupling
$f_{m_{11}m_{12}m_{13}}f_{m_{21}m_{22}m_{23}}f_{m_{31}m_{32}m_{33}}$
as a product of three factors (one for each column of
the matrix $m_{jk}$) in terms of $\ch_0$ and $\ch_1$.  The results
are tabulated in Table~\tabref{conditions.cf}.
\vskip20pt
\vbox{
$$
\vbox{\offinterlineskip\halign{
&\vrule#&\strut \quad #\quad\hfil
&\vrule#&\quad\hfil #\hfil\quad&\vrule#\cr
\noalign{\hrule}
height 2pt&\omit&&\omit&\cr
height 12pt depth 5pt&\hfil Condition&&Factor&\cr
height 2pt&\omit&&\omit&\cr
\noalign{\hrule\vskip3pt\hrule}
height 2pt&\omit&&\omit&\cr
height 12pt depth 5pt&If all the $m_{jk}\, ,~j=1,2,3,$ are distinct&&
$\;\,r_k \nu |\eta^2(\t_k) \ch_1(\t_k)|$&\cr
height 12pt depth 5pt&If precisely two of the $m_{jk}$ are distinct&&0&\cr
height 12pt depth 5pt&If all three are equal&&$
\;\,r_k \nu |\eta^2(\t_k) \ch_0(\t_k)|$&\cr
height 5pt&\omit&&\omit&\cr
\noalign{\hrule}}}
$$
\nobreak\tablecaption{conditions.cf}{The factor for the Yukawa coupling
in terms of $\ch_0$ and~$\ch_1$.}
\vskip10pt }

Evaluating the couplings in the limit $s \rightarrow 0$
corresponds to calculating the fully corrected couplings,
including instanton effects, in the orbifold limit of the
\Z--manifold.  This was first done in\ \cite\HV.
Later, in \cite\LMN\  and \cite\CMLN,
the normalized rules were written for a particular basis.
The Yukawa couplings we find are the absolute values of those
in~\cite\LMN\Footnote{\cite\CMLN\ miss a factor of $\sqrt2$
in $\imag \, \t$ when they quote the results of \cite\LMN.};
we may conclude that our results agree with previous calculations
in the orbifold limit.  Note that the ratio of the third row to the
first row is
$$
{\ch_0(\t_k) \over \ch_1(\t_k)} = \ph_k~,$$
which agrees with our calculations using the cohomology ring,
{\it cf.} Table~\tabref{conditions.Z}.
Also we see that this is indeed a sum over instantons and that without
these corrections the coupling vanishes.
As an example, we find that
$$\eqalign{
f_{111}f_{122}f_{133} &= \ph_1 e_1 e_2 e_3 \cr
                &= r_1\, \n\,\eta^2(\t_1)\, \ch_0(\t_1) \cr
                &= r_1\, \n \sum_{m,n}e^{2\p i(m^2 + mn + n^2)\t_1}~.\cr}$$

Our purpose has been to rediscover the instanton corrections to
the Yukawa couplings on the \Z\ orbifold.
However, our method may be used to do more.
Since the periods have been found as an expansion in $s$,
it is possible to examine the expansion of the Yukawa couplings
in the flat coordinate $\rho$.
This should follow the predictions of \
\cite\FLTtwo\ and \cite\Villa\
which were based on
the behaviour of the Yukawa couplings under modular transformations.
One may also consider the more general case of allowing
all $s_{mnp}$ to be nonzero.  This would allow a more
complete treatment of the modular group.
\newpage
\chapno=-1
\section{appendix1}{The Periods Calculated from a Homology Basis}
We wish to show in this Appendix a method of calculating the periods
by directly evaluating the integral of $\O$ over a homology basis.
We do this firstly in the orbifold limit, and then in the
case where one $s_{mnp}$ is nonzero.
\subsection{The Periods on the Mirror when All
\hbox{\elevenboldmath \char '163 \lower.5ex\hbox{\eightboldmath \char '155
 \eightboldmath \char '156 \eightboldmath \char '160}
  \hskip 2pt\elevenboldrm \char '075\hskip 2pt \elevenboldrm 0}}
\noindent
Recall that
$$
\O_{5,2} = {1\over{2 \p i}}\int_{C_p}{\mu \over p^3}~,$$
where $C_p$ is a circle around $p=0$,
and that in the limit where all $s_{mnp}\rightarrow 0$, we have
$$
p=\sum _{i=1}^9 y_{i}^3 - 3\sum _{k=1}^3 \ph_k e_k~.$$
We can create a  7--cycle
$$\eqalign{
\G=\{y_9&=1;\ y_8 \hbox{ given by the branch of }p=0\cr
&\hbox{ for which arg} (y_8) \rightarrow {\pi\over3}
\hbox{ as } \phi_1,\phi_2,\phi_3 \rightarrow 0; \cr
&\hskip20pt \g_1 \times\ldots\times\g_7 \}\cr}$$
using one-cycles $\g_j$ that are defined in
the $y_j$ plane:  the cycle comes in from infinity along the line
$y_j = t_j e^{2 \pi i/3}$, where $t_j$ is a real number, and goes out
along the real axis.
Along this cycle the branch choice is unique.
$C_p$ is a circle around $p=0$ in the $y_8$ plane;
this can be deformed to $\g_8$.
Now we can compute the period in the approximation that the
$\ph_i$'s are all small.  In that case we can write
$$\eqalign{
q&=\int_\G \O_{5,2}\cr
 &={1\over{2 \p i}} \int_{\g_1\times\ldots \times \g_8}
                    {{dy_1 \ldots dy_8} \over
          {(\sum_{i=1}^9 y_i^3)^3}} \sum_{n=0}^\infty {{(n+1)(n+2)}\over2}
          \left({{3 \ph_ke_k}\over{\sum_{i=1}^9 y_i^3}}\right)^n~.\cr}$$
The integrals are of the form
$$\eqalign{
I(k_1,\ldots,k_8)&=\int_{\g_1\times\ldots \times \g_8} {\prod_{i=1}^8
     y_i^{k_i} dy_i \over \left(1+\sum_{i=1}^8 y_i^3\right)^{3+n}}\cr
     &={1\over 3^8}\left(1-\o^{r_1+1}\right)\left(1-\o^{r_2+1}\right)
      {\prod_{i=1}^3 \left(1-\o^{r_i+1}\right)^2\G^3\left({r_i+1\over 3}\right)
     \over \G\left(n+3\right)}~,\cr}$$
where $r_1=k_1=k_2=k_3, r_2=k_4=k_5=k_6, r_3=k_7=k_8=k_9.$
The last equality follows by noting that
$\sum_{i=1}^9 k_i =3 n$ or $\sum_{i=1}^3 r_i = n.$
Using this we can write
$$\eqalign{
q = {1\over{2 \cdot 3^8 \cdot 2 \p i }}\sum_{r_1,r_2,r_3=0}^\infty
      &\left(1-\o^{r_1+1}\right) \left(1-\o^{r_2+1}\right) \cr
&\prod_{j=1}^3 \left\{      \left(1-\o^{r_j+1}\right)^2
     \G^3\left({r_j+1\over 3}\right)
 {\left(3 \ph_j\right)^{r_j}  \over r_j!} \right\}~.\cr}$$
This factors into a product of three sums, one for each $\ph_i$.
The sums for $\ph_1$ and $\ph_2$ are identical and will be
called $S_1$; the other we will call $S_2$.  So we have
$$
q={1\over{2 \cdot 3^8 \cdot 2 \p i }}\,S_1(\phi_1)S_1(\phi_2)S_2(\phi_3)~.$$
These factors can be written as linear combinations of hypergeometric
functions, as we will now show.  Consider
$$\eqalign{
S_1(\ph) = \sum_{r=0}^\infty\left(3\ph\right)^r &\left(1-\o^{r+1}\right)^3
    {\G^3\left(r+1\right)\over r!}\cr
=\sum_{k=0}^\infty \left(3\ph\right)^{3 k} &\left(1-\o\right)^3
    {\G^3\left(k+1/3\right)\over
\G\left(3k+1\right)}+\cr
&\sum_{k=0}^\infty \left(3\ph\right)^{3 k+1}
     \left(1-\o^2\right)^3
       {\G^3\left(k+2/3\right)\over \G\left(3k+2\right)}~.\cr}$$
The last line is obtained by splitting the original sum into
three sums
depending on the value of $r \, {\rm mod}\,3$.  Only two terms survive since
$\left(1-\o^{r+1}\right)^3=0$ if $r=2 \, {\rm mod} \, 3$.  Using the
multiplication formula for the gamma function
$$
\G(z) \G(z+1/3) \G(z+2/3) = 2\p \ 3^{1/2-3 z} \, \G(3 z)$$
we arrive at
$$
S_1(\ph)=3 \cdot 2 \p i
       \left(-\sum_{k=0}^\infty \ph^{3k}{\G^2\left(k+1/3\right)
   \over\G\left(k+2/3\right)k!}
+\ph \sum_{k=0}^\infty \ph^{3k}{\G^2\left(k+2/3\right)
   \over\G\left(k+4/3\right)k!}
       \right)~.$$

As was expected, these are linear combinations of the solutions
of Equation \eqref{Zhge}
$$\eqalign{
S_1(\ph) & =3 \cdot 2\p i  \left( -Z_1(\ph) + Z_2(\ph)\right) \cr
S_2(\ph) & =-2\p \sqrt 3 \o \left(Z_1(\ph) + \o Z_2(\ph)\right)~. \cr}$$

Other periods can be found by integrating over different 7--cycles.
We can alter $\G$ by changing any of the one--cycles.
If we choose a different one--cycle in the $x_{ij}$'th
coordinate plane, only the function in $\ph_i$ in $q$ changes.
For example, if we substitute $\hat\g_1$ for $\g_1$, where $\hat\g_1$ is
$\g_1$ rotated by $e^{2 \pi i/3}$, then we get
$${1\over{2\pi i}}\int_{\hat\g_1\times \g_2 \times\ldots\times\g_7\times C_p}
  {\m\over p^3}={1\over{2\cdot 3^8\cdot 2\p i}}
  \hat{S}_1(\phi_1)S_1(\phi_2)S_2(\phi_3)~,$$
where
$$\hat{S}_1(\phi_1) = \o S_1(\o \phi_1) = 3 \cdot 2\p i \o
   \left( -Z_1(\ph_1) + \o Z_2(\ph_1) \right)~.$$
Similarly if instead of $\g_1$, we take $\check\g_1$
defined by a rotation of $e^{4\pi i/3}$,
the factor in $q$ involving  $\phi_1$ becomes
$$\check{S}_1(\phi_1) = \o^2 S_1(\o^2 \phi_1) =
  3 \cdot 2\p i \o^2 \big( -Z_1(\phi_1) +
  \o^2 Z_2(\phi_1)\big)~.$$
Were we to have changed the $\g_7$ to $\hat \g_7$ or $\check \g_7$,
we would change the $\phi_3$ dependence to
$$\hat{S}_2(\phi_3) = \o S_2(\o \phi_3) ~\hbox{ or }~
   \check{S}_2(\phi_3) = \o^2 S_2(\o^2 \phi_3)~,$$
respectively.
Choosing a different branch for the solution of $p=0$ in the $y_8$ plane
also changes the period; $C_p$ can be deformed to $\hat\g_8$ if we take
arg$(y_8) \rightarrow \pi$, and to $\check\g_8$ if
arg$(y_8) \rightarrow {5\pi\over3}$.

We have a wealth of possible seven-cycles
using the three one-cycles $\g$, $\hat \g$, and $\check \g$.
However, this does not lead to an overabundance of independent periods.
First of all, a simple permutation of two coordinates within the
same $e_i$ has no effect: the period derived from choosing
$\hat\g_1\times\g_2 $ is the same as that from choosing
$\g_1\times\hat\g_2 $, for example.
Furthermore, since  $\g + \hat\g + \check\g = 0$,
There are many other simplifying relations that are easy to find.
As one more example, choosing $\hat\g_1\times\hat\g_2 $ gives the same period
as choosing $\check\g_1\times\g_2$.
Of course, since all these hypergeometric functions are solutions of the
same second order differential equation, each can be written as a linear
combination of only 2 independent functions.
We note that
$$\eqalign{
\hat{S}_1(\phi) &= -2 S_1(\ph) + 3 S_2(\ph)\cr
\check{S}_1(\phi)&= \hphantom{-2} S_1(\ph)  - 3S_2(\ph)\cr
\hat{S}_2(\phi) &= -\hphantom{2}S_1(\ph) + \hphantom{3}S_2(\ph)\cr
\check{S}_2(\phi) &= \hphantom{-2}S_1(\ph) -2 S_2(\ph)\cr}
\eqlabel{Zhgfrelations}$$
Since there are two linearly independent
solutions of Equation \eqref{Zhge} for each $\phi_i$,
and since $2^3 = 8$,
we conclude that we can choose to write the eight independent periods
on the manifold as products of these solutions.
\newpage
\subsection{The Expansion for One
\hbox{\elevenboldmath \char '163 \lower.5ex\hbox{\eightboldmath \char '155
 \eightboldmath \char '156 \eightboldmath \char '160}} Nonzero}
We wish to examine the periods on the manifold to some order in the $s_{mnp}$.
Suppose we let only one $s_{mnp} \equiv s$ be non-zero.
We take $$ p=\sum _{i=1}^9 y_{i}^3 - 3\sum _{k=1}^3 \ph_k e_k
    - 3  s f_{mnp}~,$$
and a variation of the seven--cycle
$$\eqalign{
\G=\{ y_9 &=1;\ y_8 \hbox{ is a solution of }p=0,\cr
&\hbox{ branch chosen by arg} (y_8) \rightarrow {\pi\over3} \hbox{ as }
\phi_1,\phi_2,\phi_3,\hbox{ and }s \rightarrow 0;\cr
&\hskip15pt \g_1 \times\ldots\times\g_7 \}~.\cr}$$
The 7--cycle is built out of one--cycles $\g_i$ and a branch choice.
As explained previously,
we could rotate some of the one--cycles by $e^{2 \pi i /3}$
to $\hat{\g}_i$, or choose a different branch.  We use the notation
$(\d_1,\d_2,\ldots,\d_8)$ to stand for the choices made:
$\d_i$=0 means that we integrate $y_i$ over $\g_i$;
$\d_i$=1 means that we integrate $y_i$ over $\hat \g_i$.
We calculate the  period via
$q=\int_{(\d_1,\d_2,\ldots,\d_8)} \O_{5,2}$,
and expand in powers of $s$, using the notation that
$q_i$ is of order $i$ in $s$:
$q =q_0 + q_1 + q_2 + \ldots$.
We further split the period up according to the value of
$r=k$ mod 3, by letting $k=3 d +r$.

It turns out that $q$ depends only on four phases
$$\eqalign{
\D_1 &= \d_1 + \d_2 +\d_3 \cr
\D_2 &= \d_4 + \d_5 +\d_6 \cr
\D_3 &= \d_7 +\d_8 \cr
\D_s &= \d_m + \d_{n+3} +\d_{p+6}~. \cr}$$
Since we have set $y_9=1$,
we take the nonzero $s_{mnp}$ to be $s_{mn1}$ or $s_{mn2}$;
then we can write the period in the following way:
$$\eqalign{
{2 \cdot 2\pi i \cdot 3^8\over (2 \pi \sqrt{3})^3} q
= & \o^{\D_1 + \D_2 + \D_3} \sum_{d=0}^\infty {(3s)^{3d} \over (3d)!} F_1\cr
    &+\o^{\D_1 + \D_2 + \D_3 + \D_s}
      \sum_{d=0}^\infty {(3s)^{3d+1} \over (3d+1)!} F_2 \cr
    &+\o^{2(\D_1 + \D_2 + \D_3 + \D_s)}
      \sum_{d=0}^\infty {(3s)^{3d+2} \over (3d+2)!} F_3\cr}
       \eqlabel{Zperiodones}$$
where
$$\eqalign{
F_1 &= 3 \o [- Z_1(\phi_1,d)+ \o^{\D_1}Z_2(\phi_1,d)]\,
   [ - Z_1(\phi_2,d)+ \o^{\D_2}Z_2(\phi_2,d)]\times \cr
   &\hskip25pt [ Z_1(\phi_3,d)+ \o^{1+\D_3}Z_2(\phi_3,d)]\cropen{5pt}
F_2 &= -3 \o [ Z_3(\phi_1,d)]\,
       [ Z_3(\phi_2,d)]\,
       [ Z_3(\phi_3,d)]\cropen{5pt}
F_3 &= -3 \o^2 [ Z_4(\phi_1,d)]\,
       [ Z_4(\phi_2,d)]\,
       [ Z_4(\phi_3,d)] \cropen{5pt}}$$
and
$$\eqalign{
Z_1(\phi,d) &=
  {\G\left(\third{1}\right)\G\left(d+\third{1}\right)\over
       \G\left(\third{2}\right)}
\F(\third{1},d + \third{1};\third{2};\phi^3) \hphantom{\ph}
\quad,\quad Z_3(\phi,d) =Z_1(\phi,d+\third{1})~,\cropen{5pt}
Z_2(\phi,d) &=
  {\G\left(\third{2}\right)\G\left(d+\third{2}\right)\over
       \G\left(\third{4}\right)}\phi\,
\F(\third{2},d + \third{2};\third{4};\phi^3)
\quad,\quad Z_4(\phi,d) =Z_2(\phi,d+\third{2})
{}~. \cr} \eqlabel{Zhgfd}$$

Now if we take the combination
$$\eqalign{
&q(\D_1,\D_2,\D_3,\D_s)+ q(\D_1,\D_2,\D_3,\D_s+1)
+ q(\D_1,\D_2,\D_3,\D_s+2) \cr
&={(2 \pi \sqrt{3})^3\over 2 \cdot 2\pi i \cdot 3^7} \,
\o^{\D_1 + \D_2 + \D_3} \sum_{d=0}^\infty {(3s)^{3d} \over (3d)!} F_1\cr}
$$
we find periods that are independent of $\D_s$, but they
still depend on $\D_1$, $\D_2$ and $\D_3$ independently.
Thus there are eight of these.
They reduce to our familiar eight in the limit $s \rightarrow 0$.
The first correction to these periods is of order $s^3$.
The integral periods $z^0,\ldots,z^3,\ca{G}_0,\ldots,\ca{G}_3$ of
Equation~\eqref{Piten} are linear combinations of these.
The change of basis involves a symplectic transformation times a
scale factor, as can be inferred from \eqref{QR}.

If we form the combination
$$\eqalign{
&q(\D_1,\D_2,\D_3,\D_s)+ q(\D_1+1,\D_2,\D_3,\D_s+2)
+ q(\D_1+2,\D_2,\D_3,\D_s+1) \cr
&= {(2 \pi \sqrt{3})^3\over 2 \cdot 2\pi i \cdot 3^7} \Big\{ \,
  \o^{\D_1 + \D_2 + \D_3 + \D_s}
      \sum_{d=0}^\infty {(3s)^{3d+1} \over (3d+1)!} F_2 \cr
  &\hskip70pt +\o^{2(\D_1 + \D_2 + \D_3 + \D_s)}
      \sum_{d=0}^\infty {(3s)^{3d+2} \over (3d+2)!} F_3\Big\}\cr}
$$
we find that there are only two new independent
periods, as there should be.  (Recall that the number of periods is
$2(b_{4,3}+1)$; so for each nonzero (4,3)--form there are two
periods.)
Choosing the periods formed when $\D_1 + \D_2 + \D_3 + \D_s = 1$ and $2$
and applying the same change of scale as for the other periods leads
us to the periods $z^4$ and $\ca{G}_4$ of \eqref{G4z4} and \eqref{Manda}.

Altogether we now have 10 independent periods;
the eight we found when all $s_{mnp}=0$ get corrected by
terms of order $s^{3d}$
and the two new ones have terms of order $s^{3d+1}$ and $s^{3d+2}$.
With foresight we could have chosen the gauge
$$ \O_{5,2} = 6 (2 \pi i)^3 f \int {\m \over p^3} $$
using the factor $f$ as given in \eqref{Zfchoice} and incorporating
the change of scale mentioned above.
The integral periods of \eqref{Piten} can be calculated
directly by integrating this over
a homology basis derived from the cycles defined above.

As a check on the relative normalization of the integral periods,
we compute the value of
$$W_s =  \int \O \wedge \partial_s \O =
\ca{G}_a \partial_s z^a - z^a \partial_s \ca{G}_a ~.$$
This should vanish, since $\O \in H^{5,2}$ and
$ \partial_s \O \in H^{5,2} \oplus H^{4,3} $.
Looking at the expansion to $\ca{O}(s^2)$,
$$W_s = {3^3 s^2 \over 2} f^2 \Big\{
          \prod_j \big( R(\ph_j,0) Q(\ph_j,1) - Q(\ph_j,0) R(\ph_j,1) \big)
           + (\o^2 - \o) c^2 \prod_j Z_3(\ph_j) Z_4(\ph_j) \Big\} ~,$$
so using
$$ \eqalign{ R(\ph,0) Q(\ph,1) - Q(\ph,0) R(\ph,1) &=
{ (2 \pi)^5 i \over 3^2} \ph (1-\ph^3)^{-1}\cr
 Z_3(\ph) Z_4(\ph) &= {2 \pi \over \sqrt3}  \ph (1-\ph^3)^{-1} \cr}$$
we find that $W_s = 0$ only when $c = {(2 \pi)^6 i \over 3^{5/2}}$.

\newpage
\acknowledgements
It is a pleasure to thank
Paul Aspinwall, Xenia de la Ossa, Tristan H\"ubsch, Anamaria Font, Jan Louis,
Dieter L\"ust, Fernando Quevedo, Cumrun Vafa and Nick Warner
for instructive discussions. We are indebted to Vadim Kaplunovsky for making
computer resources available.
\vskip0.5in
\immediate\closeout\referencewrite
\referenceopenfalse\rightskip0pt plus4em
\line{\bf\hfil References\hfil}\bigskip\parskip=5pt
\input referenc.texauxil
\bye